\RequirePackage{lineno}

\documentclass[preprint,12pt]{revtex4}

\usepackage{epsfig}
\usepackage{amsmath,amssymb,bm}
\usepackage{hyperref}
\usepackage{placeins}
\usepackage{pstricks}

\def\lsim{\mathrel{\rlap{\lower4pt\hbox{\hskip1pt$\sim$}}
    \raise1pt\hbox{$<$}}}         
\def\gsim{\mathrel{\rlap{\lower4pt\hbox{\hskip1pt$\sim$}}
    \raise1pt\hbox{$>$}}}         
    
\newcommand*{\TeV}{\ensuremath{\text{Te\kern -0.1em V}}}
\newcommand*{\GeV}{\ensuremath{\text{Ge\kern -0.1em V}}}
\newcommand*{\MeV}{\ensuremath{\text{Me\kern -0.1em V}}}
\newcommand*{\keV}{\ensuremath{\text{ke\kern -0.1em V}}}
\newcommand*{\eV}{\ensuremath{\text{e\kern -0.1em V}}}

\usepackage{blindtext}

\begin{document}

\preprint{DESY 19-010}
\title{Probing the photonic content of the proton using photon-induced dilepton production in $p+\textrm{Pb}$ collisions at the LHC}

\author{M. Dyndal}
\email{mateusz.dyndal@cern.ch}
\address{Deutsches Elektronen-Synchrotron DESY, Hamburg, Germany}

\author{A. Glazov}
\email{alexander.glazov@desy.de}
\address{Deutsches Elektronen-Synchrotron DESY, Hamburg, Germany}

\author{M. Luszczak}
\email{luszczak@ur.edu.pl}
\address{Faculty of Mathematics and Natural Sciences, University of Rzeszow, Poland}
\author{R. Sadykov}
\email{renat.sadykov@cern.ch}
\address{Joint Institute for Nuclear Research (JINR), Dubna, Russia}

\begin{abstract}
We propose a new experimental method to probe the photon parton distribution function inside the proton (photon PDF) at LHC energies.
The method is based on the measurement of dilepton production from the $\gamma p\rightarrow\ell^+\ell^-+X$ reaction in proton--lead collisions. These experimental conditions guarantee a clean environment, both in terms of reconstruction of the final state and in terms of possible background.
We firstly calculate the cross sections for this process with collinear photon PDFs, where we identify optimal choice of the scale, in analogy to deep inelastic scattering kinematics.
We then perform calculations including the transverse-momentum dependence of the probed photon.
Finally we estimate rates of the process for the existing LHC data samples.
\end{abstract}


\maketitle

\section{Introduction}

Precise calculations of various electroweak reactions in $pp$ collisions at the LHC need to account for, on top of the higher-order corrections, the effects of photon-induced processes.
The relevant examples are the production of lepton pairs~\cite{Aad:2014qja, Aad:2016zzw,Accomando:2016tah, Luszczak:2015aoa, Harland-Lang:2016apc} or pairs of electroweak bosons~\cite{Luszczak:2014mta, Denner:2015fca, Dyndal:2015hrp, Ababekri:2016kkj, Biedermann:2016guo, Biedermann:2016yvs, Yong:2016njr, Luszczak:2018ntp}.

Recently, a precise photon distribution inside the proton has been evaluated in Ref.~\cite{Manohar:2016nzj}.
This approach provides a model-independent determination of the photon PDF (embedded in the so-called LUXqed distribution)
and  it is based on proton structure function and elastic form factor fits in electron--proton scattering.

To date, there are no experimentally clean processes identified that would allow verification or strong constraint of the calculations.
For example, the extraction of the photon PDF from isolated photon production in deep inelastic scattering (DIS)~\cite{Schmidt:2015zda} 
or from inclusive $pp\rightarrow\ell^+\ell^-+X$~\cite{Ball:2013hta, Aad:2016zzw, Giuli:2017oii} is limited due to large QCD background.
On the contrary, the elastic part of the photon PDF is verified via exclusive $\gamma\gamma\rightarrow\ell^+\ell^-$ process, measured in $pp$ collisions by ATLAS~\cite{Aad:2015bwa,Aaboud:2017oiq}, CMS~\cite{Chatrchyan:2011ci,Chatrchyan:2012tv} and recently by CMS+TOTEM~\cite{Cms:2018het} collaborations.

\begin{figure}[h!]
\includegraphics[width=1.\textwidth]{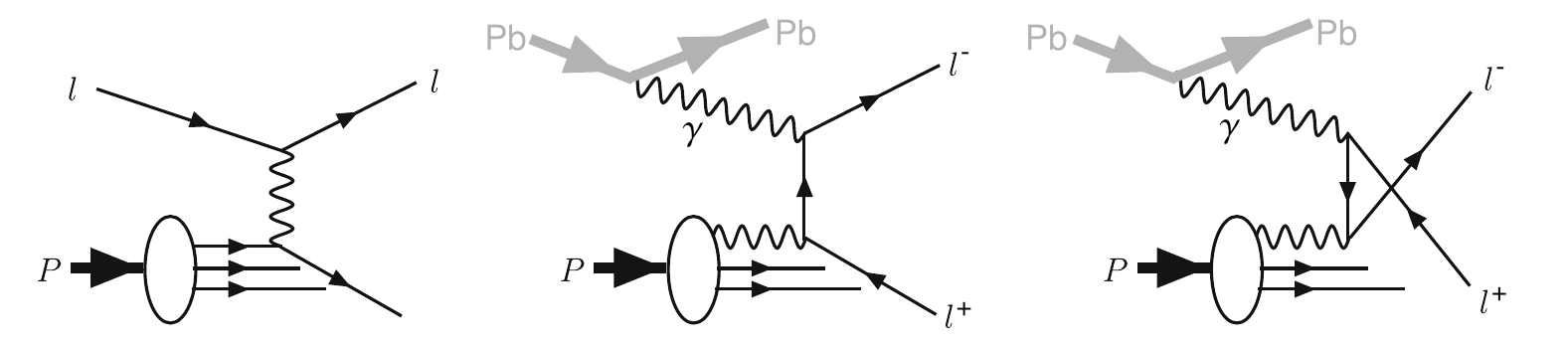}
 \put(-400,-3){{\footnotesize(a)}}
 \put(-246,-3){{\footnotesize(b)}}
\put(-90,-3){{\footnotesize(c)}}
\caption{Schematic graphs for deep inelastic scattering, $\ell^{\pm} p\rightarrow \ell^{\pm} +X$ (a) and photon-induced dilepton prodcution, $\gamma p\rightarrow \ell^+\ell^- + X$, in $p+\textrm{Pb}$ collisions for $t$-channel (b) and $u$-channel (c) lepton exchange.}
\label{fig:diagrams}
\end{figure}

We therefore propose a new experimental method to constrain the photonic content of the proton.
Due to the large fluxes of quasi-real photons from the lead ion (Pb) at the LHC, the photon-induced dilepton production in $p+\textrm{Pb}$ collision configuration (where Pb serves as a source of elastic photons) is a very clean way to probe the photon PDF inside a proton. 
This process is shown schematically in Fig.~\ref{fig:diagrams}, where by analogy to DIS, two leading-order diagrams can be identified.
Since the photon flux from the ion scales with $Z^2$ ($Z$ is the charge of the ion) and QCD-induced cross-sections scale approximately with the atomic number $A$,
the amount of QCD background is greatly reduced comparing to the $pp$ case.

Moreover, as this process does not involve the exchange of color with the photon-emitting nucleus, no significant particle production is expected in the rapidity region between the dilepton system and the nucleus. 
The photon-emitting nucleus is also expected to produce no neutrons because the photons couple to the entire nucleus. 
Thus a combination of requirements on rapidity gap and zero neutrons in the same direction provide straightforward criteria to identify these events experimentally. 

\section{Formalism}

\subsection{Elastic photon fluxes}

To get the distribution of the elastic photons from the proton, one can express the equivalent photon flux through
the electric and magnetic form factors $G_E(Q^2)$ and $G_M(Q^2)$ of the proton.
This contribution is obtained as
\begin{eqnarray}
   \gamma^{p}_{el}(x,Q^2) = \frac{\alpha_{\rm{em}}}{\pi}
\Big[ \Big( 1- {x \over 2} \Big)^2 \, {4 m_p^2 G_E^2(Q^2) + Q^2 G_M^2(Q^2) \over 4m_p^2 + Q^2} + {x^2 \over 4} G_M^2(Q^2) \Big]~,
\label{proton_el_flux}
\end{eqnarray}
where $x$ is the momentum fraction of the proton taken by the photon, $Q^2$ is the photon virtuality, $\alpha_{\rm{em}}$ is the electromagnetic structure constant and $m_p$ is the proton mass.

To express the elastic photon flux for the nucleus ($\gamma^{\rm Pb}_{el}$), we follow Ref.~\cite{Budnev:1974de} and replace 
\begin{eqnarray}
 {4 m_p^2 G_E^2(Q^2) + Q^2 G_M^2(Q^2) \over 4m_p^2 + Q^2} \longrightarrow Z^2 F_{\rm em}^2(Q^2)~,
 \end{eqnarray}
where $F_{\rm em}(Q^2)$ is the electromagnetic form factor of the nucleus and $Z$ is its charge.
We also neglect the magnetic form factor of the ion in the following.

For the Pb nucleus, we use the form factor parameterization from the STARlight MC generator~\cite{Klein:2016yzr}:
\begin{eqnarray}
 F_{\rm em}(Q^2) = {3 \over (QR_A)^3}\Big[ \sin(QR_A) - QR_A \cos(QR_A) \Big] { 1 \over 1 + a^2 Q^2}~,
 \label{pb_el_flux}
\end{eqnarray}
where $R_A = 1.1 A^{1/3}$ fm, $a = 0.7$ fm and $Q = \sqrt{Q^2}$.

The elastic photon PDFs of the proton and lead nucleus can be integrated over $Q^2$ as 
\begin{equation}
\gamma^{(p,Pb)}_{el}(x)  = \int d Q^2 \gamma^{(p,Pb)}_{el}(x, Q^2) \,.
 \label{integral_flux}
\end{equation}
This is useful for the collinear-factorization approach since the $Q^2$ dependence factorizes in this case. 
\subsection{Collinear-factorization approach and choice of the scale}

The inelastic processes, with breakup of a proton, can be also considered.
At LO and at a given scale $\mu^2$, the photon parton distribution $\gamma^p_{inel}(x,\mu^2)$ of photons carrying a fraction $x$ of the proton's momentum, obeys the DGLAP equation:
\begin{eqnarray}
{d \gamma^p_{inel}(x,\mu^2) \over d \log \mu^2} =&& {\alpha_{\rm{em}} \over 2 \pi} \int_x^1 {dy \over y} 
\Big [ \sum_q P_{\gamma \leftarrow q}(y) 
 q ({x \over y}, \mu^2 )   + P_{\gamma \leftarrow \gamma}(y) \gamma^p_{inel}({x \over y},\mu^2) \Big ]~,
\end{eqnarray}
where $q (x,\mu^2)$ is the quark PDF,  $P_{\gamma \leftarrow q}$ is the $q\rightarrow\gamma$ splitting function, and $P_{\gamma \leftarrow \gamma}$ corresponds to the virtual self-energy correction to the photon propagator.
This is the basis for colinear photon-PDFs in the initial~\cite{Gluck:2002fi, Martin:2004dh} and more recent~\cite{Ball:2013hta, Martin:2014nqa, Schmidt:2014aba, Harland-Lang:2016kog, Giuli:2017oii, Manohar:2016nzj, Bertone:2017bme} analyses.

The computation of the photon-induced dilepton production cross section  requires definition of the  scale ($\mu^2$) at which the photon PDFs are convoluted.
The usual choice for $\mu$ is the mass of the system (motivated by the $s$-channel quark--antiquark annihilation process) or the transverse momentum of the leading object. 
These choices are however not optimal for the $t$- and $u$-channel initiated photon-induced process.
By analogy to DIS (Fig.~\ref{fig:diagrams}), where the scale is associated with the virtuality of the exchanged photon,
it is possible to define the scale in case of the $\gamma\gamma\rightarrow\ell^+\ell^-$ process.
This is achieved by taking the virtuality of the massive $t$- or $u$-channel propagator (Fig.~\ref{fig:diagrams}b or c).
Hence, $\mu^2 = -(p^{\gamma^{Pb}}-p^{\ell^-})^2$ for the $t$-channel diagram and $\mu^2 = -(p^{\gamma^{Pb}}-p^{\ell^+})^2$ for the $u$-channel exchange, where $p^{\gamma^{Pb}}$ is the four momentum
of the photon emitted by lead and $p^{\ell^{\pm}}$ is the four momentum of the lepton of the corresponding charge.
Note that the $u$- and $t$- channel diagrams have vanishing interference in the zero lepton mass limit. Therefore, they can be separated  while convoluting PDFs with the partonic cross section.

In the collinear approach, the $p+\textrm{Pb}\rightarrow \textrm{Pb} + \ell^+\ell^- + X$ production cross section can be written as
\begin{equation}
\sigma 
= S^2 \int dx_p dx_{\rm Pb} 
\Big [ \gamma^{p}_{el}(x_p) + \gamma^{p}_{inel}(x_p,\mu^2) \Big]
 \gamma^{\rm Pb}_{el}(x_{\rm Pb})
\sigma_{\gamma \gamma \rightarrow \ell^+ \ell^-}(x_p, x_{\rm Pb}) \,,
\label{collinear_factorization_formula}
\end{equation}
where $\sigma_{\gamma \gamma \rightarrow \ell^+ \ell^-}$
is the elementary cross section for the $\gamma \gamma \rightarrow \ell^+ \ell^-$ subprocess and $S^2$ is the so-called survival factor which takes into account the requirement that there be no hadronic interactions between the proton and the ion.

\subsection{$k_T$-factorization approach}

At lowest order, the calculations with collinear photons  produce leptons that are back-to-back in transverse kinematics.
The transverse momentum appears at higher orders, however to describe full transverse momentum spectrum  one needs  
to match the calculations to resummation or dedicated parton shower algorithms. This approach is not considered in this paper.

In the $k_T$ factorization approach (also named as high-energy factorization), one can parametrize the $\gamma^*p \rightarrow X$ vertices in terms of the proton structure functions. The photons from inelastic production have transverse momenta and non-zero virtualities $Q^2$ and the unintegrated photon distributions are used, in contrast to collinear distributions.
In the DIS limit, the unintegrated inelastic photon flux can be obtained using the following equation~\cite{daSilveira:2014jla, Luszczak:2015aoa}:

\begin{eqnarray}
\gamma^p_{inel}(x,Q^2) = {1\over x} 
{1 \over \pi Q^2} \, \int_{M^2_{\rm thr}} dM_X^2 {\cal{F}}^{\mathrm{in}}_{\gamma^* \leftarrow p} (x,\vec{q}_T^2,M^2_X) \, ,
\end{eqnarray}
and we use the functions $ {\cal{F}}^{\mathrm{in}}_{\gamma^* \leftarrow p}$ from \cite{Budnev:1974de, Luszczak:2018ntp}:
\begin{eqnarray}
{\cal{F}}^{\mathrm{in}}_{\gamma^* \leftarrow p} (x,\vec{q}_T^2, M_X) &=& {\alpha_{\rm em} \over \pi} 
\Big\{(1-x) \Big( {\vec{q}_T^2 \over \vec{q}_T^2 + x (M_X^2 - m_p^2) + x^2 m_p^2  }\Big)^2  
{F_2(x_{\rm Bj},Q^2) \over Q^2 + M_X^2 - m_p^2}  \nonumber \\
&+& {x^2 \over 4 x^2_{\rm Bj}}  
{\vec{q}_T^2 \over \vec{q}_T^2 + x (M_X^2 - m_p^2) + x^2 m_p^2  }
{2 x_{\rm Bj} F_1(x_{\rm Bj},Q^2) \over Q^2 + M_X^2 - m_p^2} \Big\} \, .
\label{eq:flux_in}
\end{eqnarray}
The virtuality $Q^2$ of the photon depends on the photon transverse momentum ($\vec{q}_T^2$) and the proton remnant mass ($M_X$):
\begin{eqnarray}
Q^2 =  {\vec{q}_T^2 + x (M_X^2 - m_p^2) + x^2 m_p^2 \over (1-x)} \, .
\label{eq:q2}
\end{eqnarray}
Moreover, the proton structure functions $F_1(x_{\rm Bj},Q^2)$ and $F_2(x_{\rm Bj},Q^2)$ require the argument
\begin{eqnarray}
x_{\rm Bj} = {Q^2 \over Q^2+M^2_X -m_p^2}.
\end{eqnarray}
Note that in Eq.~\ref{eq:flux_in} instead of using $F_2(x_{\rm Bj},Q^2),F_1(x_{\rm Bj},Q^2)$, 
we in practice use the pair $F_2(x_{\rm Bj},Q^2),F_L(x_{\rm Bj},Q^2)$, where
\begin{eqnarray}
F_L(x_{\rm Bj},Q^2) = \Big( 1 + {4 x_{\rm Bj}^2 m_p^2 \over Q^2} \Big) F_2(x_{\rm Bj},Q^2) - 2 x_{\rm Bj} F_1(x_{\rm Bj},Q^2)
\end{eqnarray}
is the longitudinal structure function of the proton.

These unintegrated photon fluxes enter the $p+\textrm{Pb}\rightarrow \textrm{Pb} + \ell^+\ell^- + X$ production cross section as

\begin{equation}
\sigma = S^2 \int dx_p dx_{\rm Pb} d\vec{q}_T \Big[ \gamma^{p}_{el}(x_p, Q^2) + \gamma^{p}_{inel}(x_p,Q^2) \Big]
 \gamma^{\rm Pb}_{el}(x_{\rm Pb})
\sigma_{\gamma^{*}  \gamma \rightarrow \ell^+ \ell^-}(x_p, x_{\rm Pb}, \vec{q_T}) \,,
\label{kt_factorization_formula}
\end{equation}
where $\sigma_{\gamma^{*} \gamma \rightarrow \ell^+ \ell^-}$ is the off-shell elementary cross-section (for details see Refs.~\cite{daSilveira:2014jla, Catani:1990eg}) and  for  $x_p \ll 1$ we have $Q^2 \approx \vec{q}_T^2$ (see Eq.~\ref{eq:q2}).
One should note that while the fluxes do not depend on the direction of $\vec{q}_T$, averaging over directions
of $\vec{q}_T$ in the off-shell cross section replaces the average over photon polarizations in the collinear case.

\section{Example experimental configuration and possible background sources}
\label{sec:experiment}

We assume collision setup from recent $p+\textrm{Pb}$ run at the LHC, carried out at the centre-of-mass energy per nucleon pair $\sqrt{s_{N N}} = 8.16$~\TeV.
Since the energy per nucleon in the proton beam is larger than in the lead beam, the nucleon--nucleon centre-of-mass system has a rapidity in the laboratory frame of $y = 0.465$.

As an example of method's applicability, we will use the geometry of ATLAS~\cite{Aad:2008zzm} and CMS~\cite{Chatrchyan:2008aa} detectors in the following.
We consider only the dimuon channel, however the integrated results for $ee$ and $\mu\mu$ channels can be obtained by simply multiplying the dimuon cross-sections by a factor of two.

We start by applying a minimum transverse momentum requirement of 4~\GeV\ to both muons.
This requirement is imposed to ensure high lepton reconstruction and triggering efficiency.
Moreover, due to limited acceptance of the detectors, each muon is required to have a pseudorapidity ($\eta^{\ell}$) that satisfies $|\eta^{\ell}|<2.4$.
Our calculations are carried out for a minimum dilepton invariant mass of $m_{\ell^+\ell^-} = 10$~\GeV. 
Such a choice is due to removal of possible contamination from $\Upsilon(\rightarrow \ell^+\ell^-)$ photoproduction.
A summary of all selection requirements is presented in Table~\ref{tab:fidRegion}.

\begin{table}[t!]
  \begin{center}
    \begin{tabular}{|l|c|}
      \hline 
    Variable  & Requirement \\ \hline
    lepton transverse momentum, $p_{\textrm{T}}^{\ell}$ & $>4$~\GeV \\
    lepton pseudorapidity, $|\eta^\ell|$ & $<2.4$ \\
    dilepton invariant mass, $m_{\ell^+\ell^-}$ & $>10$~\GeV  \\
      \hline 
    \end{tabular}
  \end{center}
  \caption{Definition of the fiducial region used in the studies.}
  \label{tab:fidRegion}
\end{table}

Possible background for this process can arise from inclusive lepton-pair production, e.g. from Drell--Yan process~\cite{Drell:1970wh,Aad:2015gta,Khachatryan:2015pzs,Alice:2016wka}.
This processes would lead to disintegration of the incoming ion, and zero-degree calorimeters (ZDC)~\cite{Dellacasa:1999ke,ATLAS:2007aa} can be used to veto very-forward-going neutral fragments which would allow this background 
to be reduced fully.
Another background can arise from diffractive interactions, hence possibly mimicking the signal topology.
However, since the Pb nucleus is a fragile object (with the nucleon binding energy of just 8 MeV) even the softest diffractive interaction will likely result in the emission of a few nucleons from the ion, detectable in the ZDC.

Another background category is the photon-induced process with a resolved photon, i.e. 
$\gamma p\rightarrow Z/\gamma^*+X$ reaction.
Here, the rapidity gap is expected to be smaller than in the signal process due to the additional particle production associated with the ``photon remnant''.
Any other residual contamination of this process can be controlled using a dedicated sample, with a dilepton invariant mass around the $Z$-boson mass.

\section{Results with collinear photon-PDFs}

We start with the calculation of the elastic contribution, $p+\textrm{Pb}\rightarrow p+\textrm{Pb}+ \ell^+\ell^-$
for which the following parameterization is used~\cite{Budnev:1974de}:
\begin{equation} \label{eq:elasticRenat}
\gamma^{p}_{el}(x)  = \frac{\alpha_{\rm em}}{\pi}
\left(
\frac{1-x+0.5x^2}{x}
\right)
\left(
\frac{F+3}{F-1}\log{F}-\frac{17}{6}-\frac{4}{3F}+\frac{1}{6F^2}
\right)~,
\end{equation}
where $F = 1+\frac{Q_0^2(1-x)}{x^2 m_p^2}$ and $Q_0^2 = 0.71$~\GeV$^2$. This parameterization is a good analytical approximation of Eq.~\ref{proton_el_flux} integrated over $Q^2$.
The results for the elastic case are cross-checked with the calculation from STARlight MC and a good agreement between the fiducial cross-sections is found:
$\sigma_{fid}^{\textrm{el}} = 17.5$~nb, whereas $\sigma_{fid}^{\textrm{STARlight}} = 17.0$~nb.
Both calculations are also corrected by a factor $S^2=0.96$ which 
is calculated using STARlight, where the hard-sphere proton--nucleus requirement~\cite{Klein:2016yzr} is used.

Next, for the inelastic case ($\gamma p\rightarrow \ell^+\ell^- + X$), several recent parameterizations of the photon parton distributions are studied: CT14qed~\cite{Schmidt:2015zda}, HKR16qed~\cite{Harland-Lang:2016kog}, LUXqed17~\cite{Manohar:2017eqh} and NNPDF3.1luxQED~\cite{Bertone:2017bme}. 
All predictions are scaled by $S^2=0.95$, again derived from STARlight. This value of $S^2$ is lower than for the purely elastic case, due to slightly smaller average impact parameter between the proton and the ion in the inelastic reaction.
One should note that all of these PDF sets include both elastic and inelastic parts of the photon spectrum.
We keep the elastic part now (as provided by each group), but we subtract it later in Sec.~\ref{sec:discussion} for the comparison with $k_T$-factorization results.

\begin{figure}[]

\includegraphics[width=0.43\textwidth]{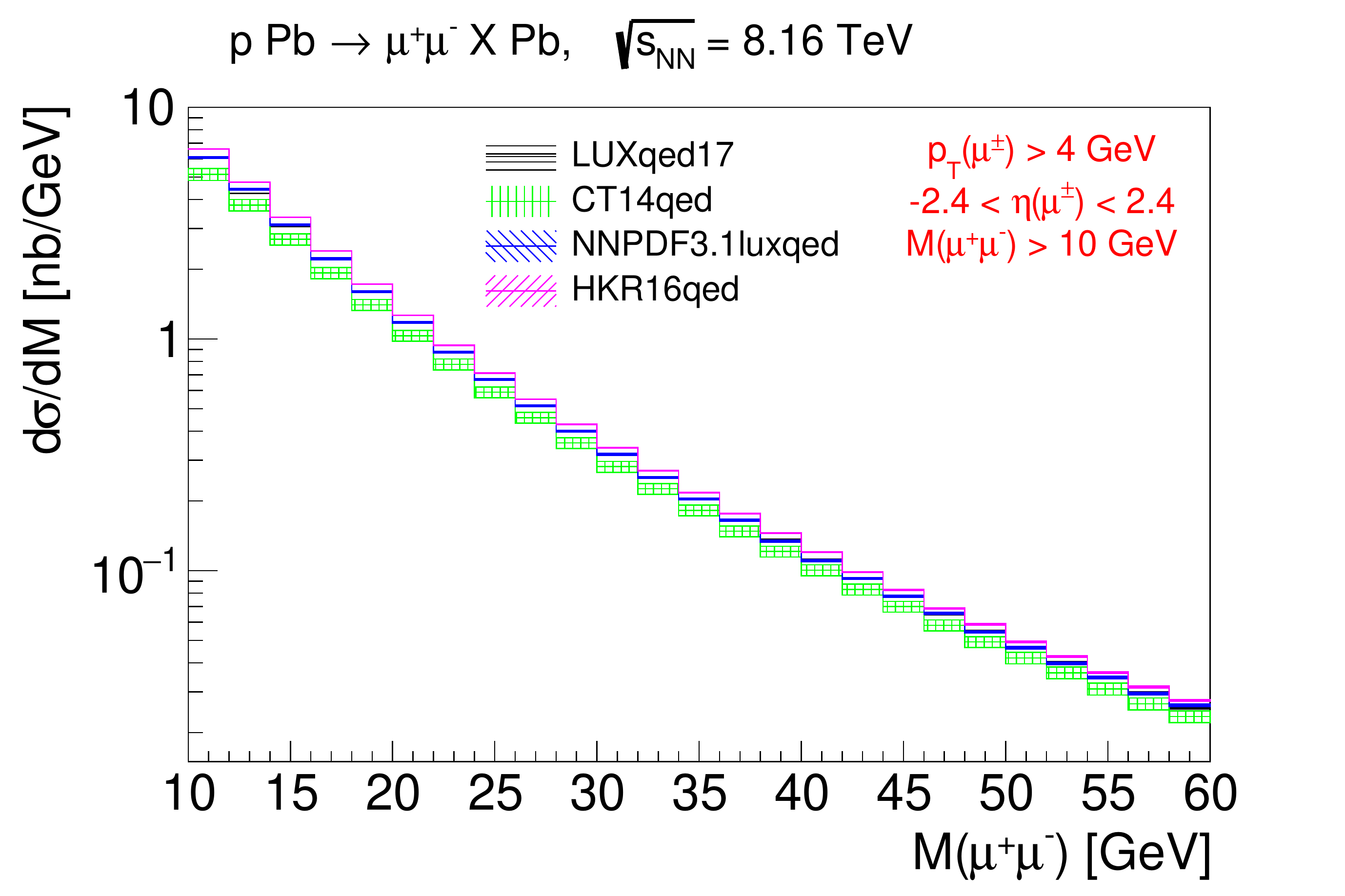}
\includegraphics[width=0.43\textwidth]{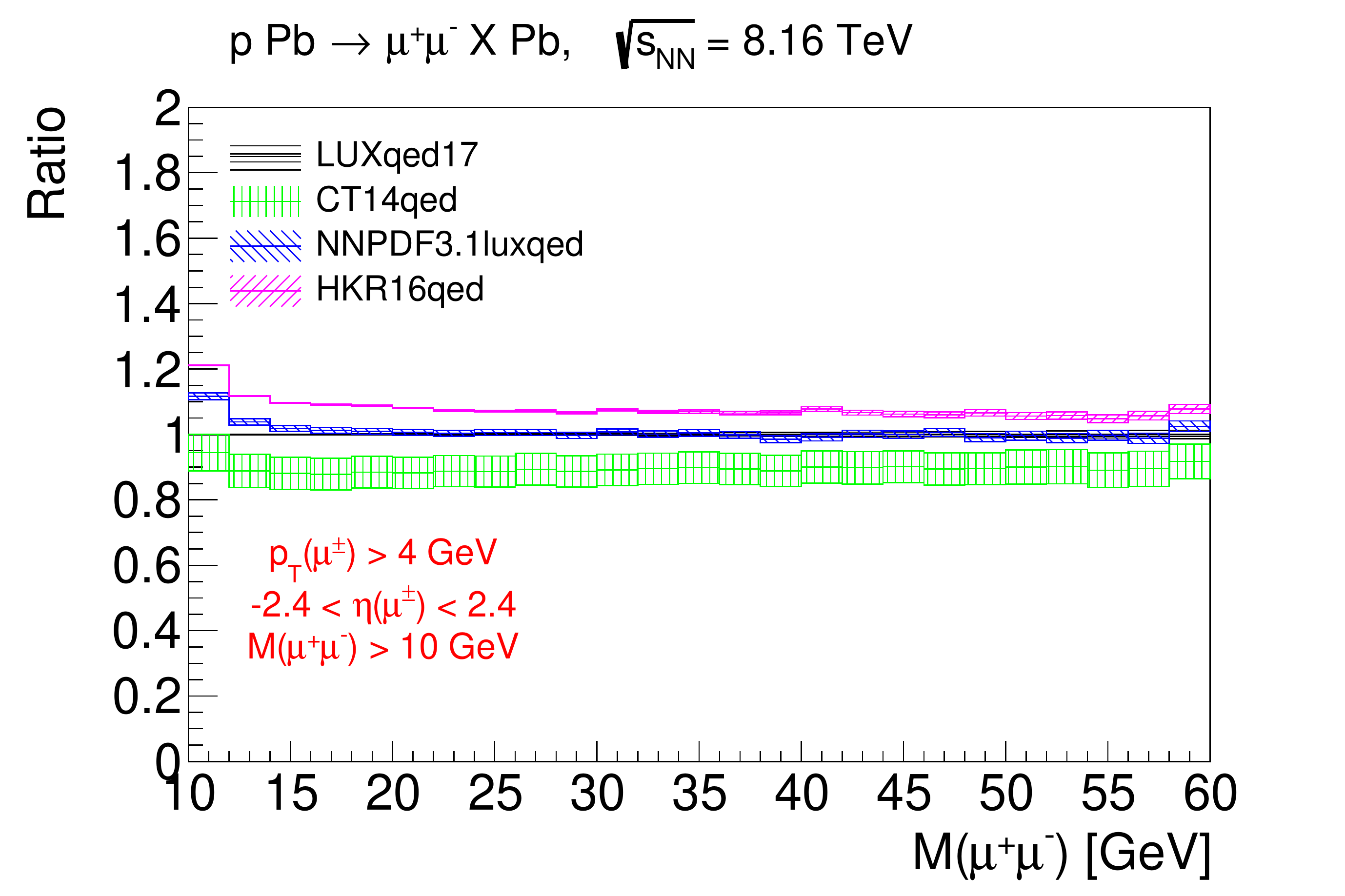}
\includegraphics[width=0.43\textwidth]{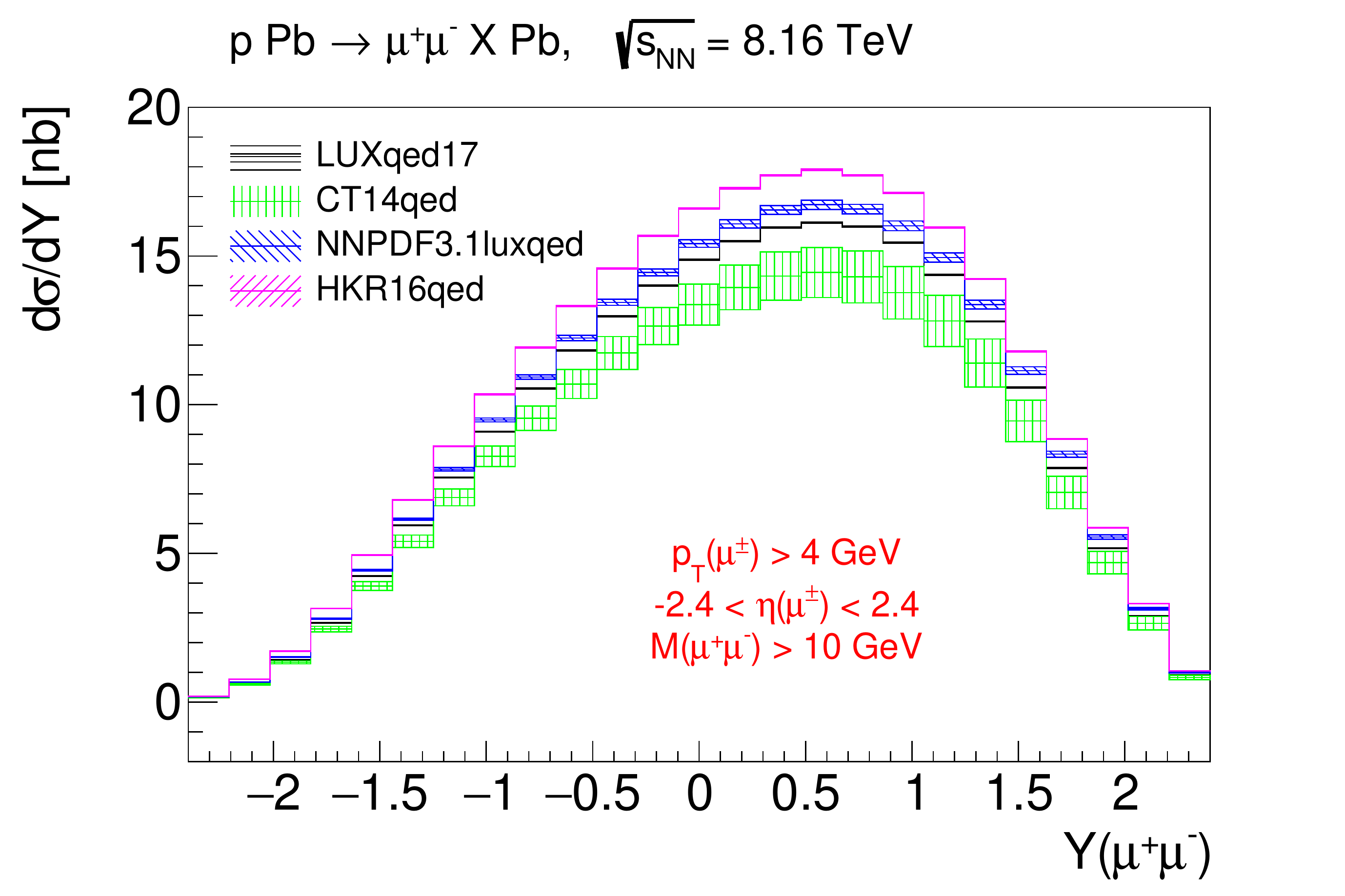}
\includegraphics[width=0.43\textwidth]{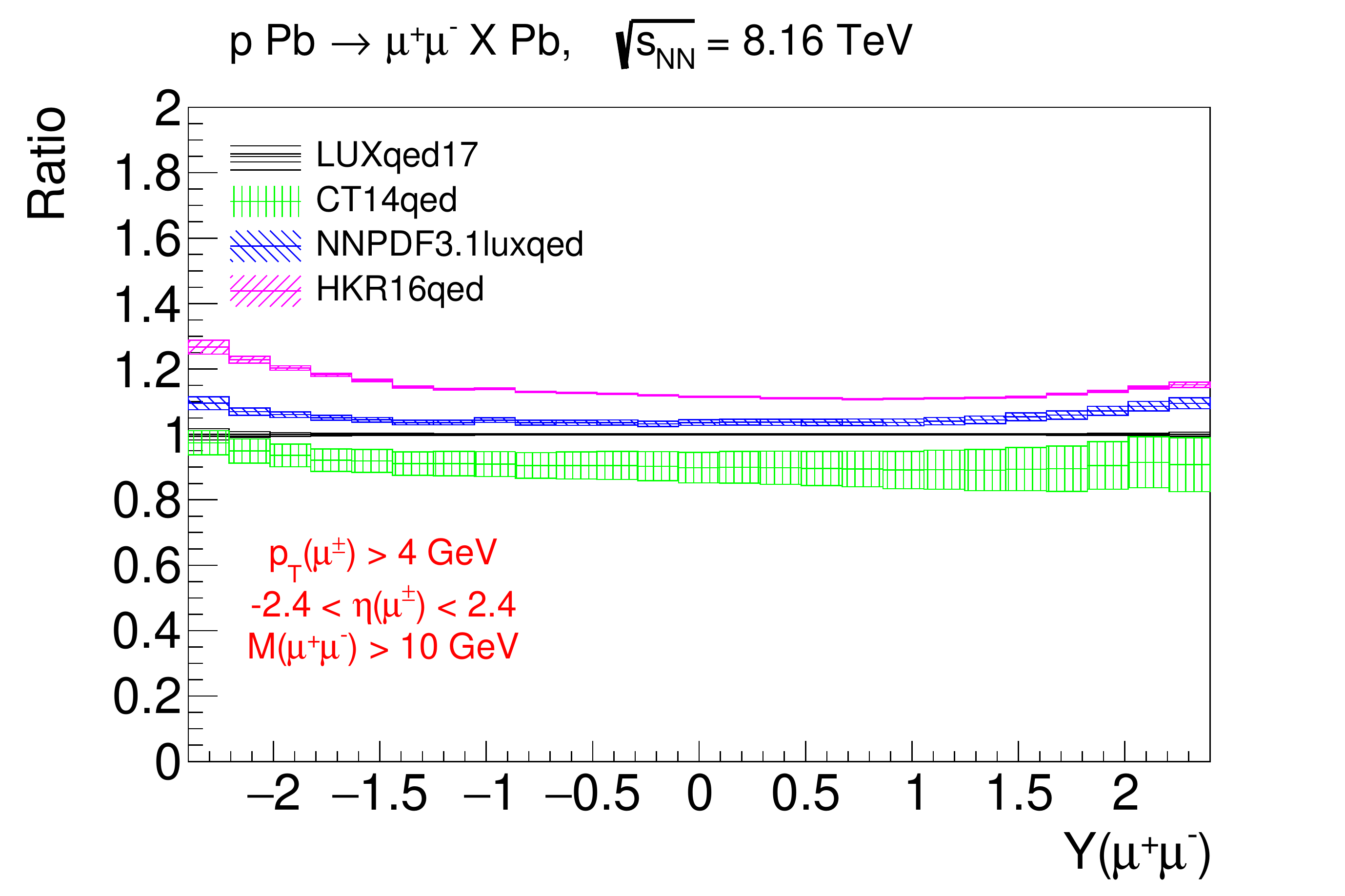}
\includegraphics[width=0.43\textwidth]{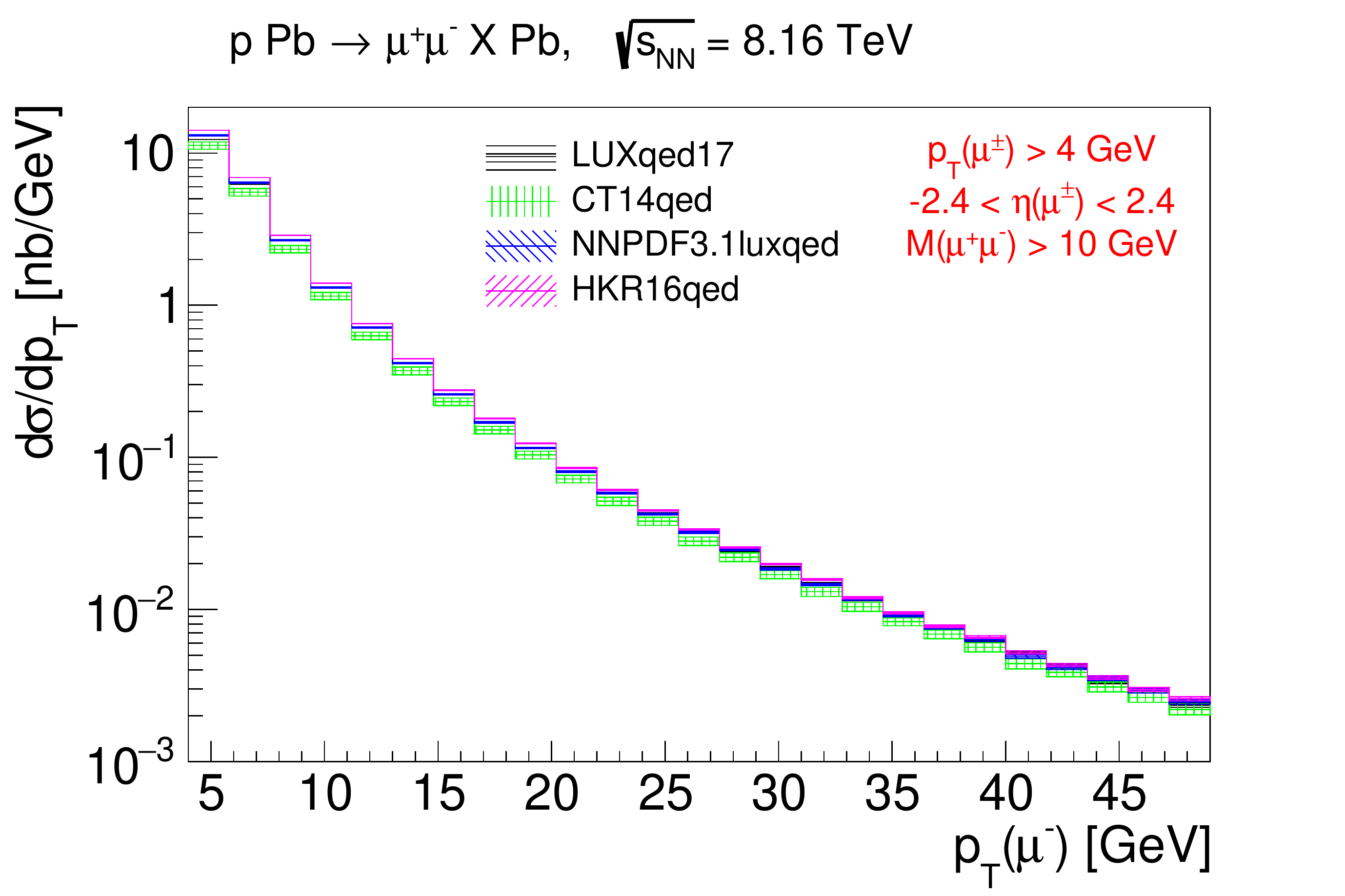}
\includegraphics[width=0.43\textwidth]{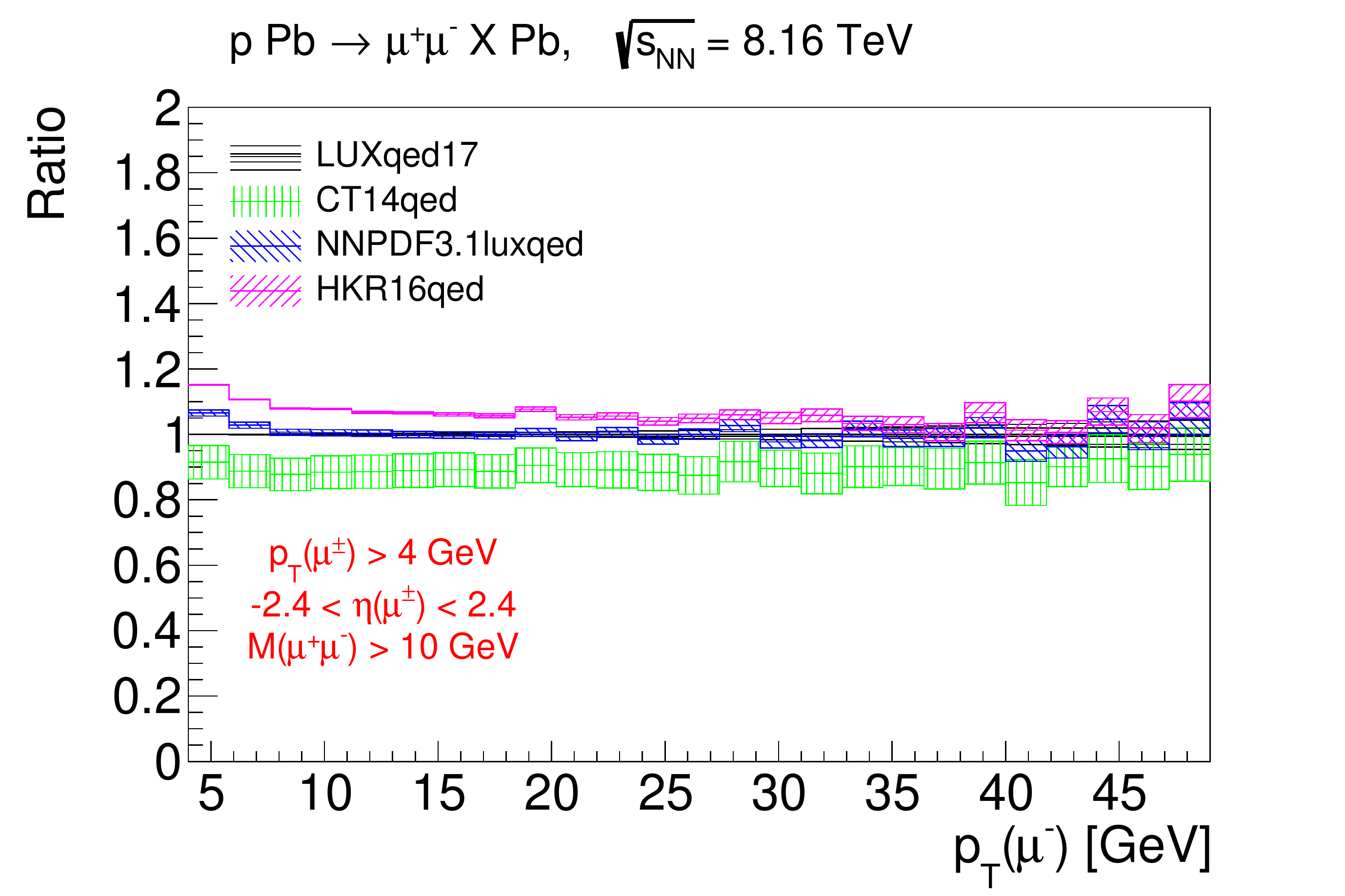}
\includegraphics[width=0.43\textwidth]{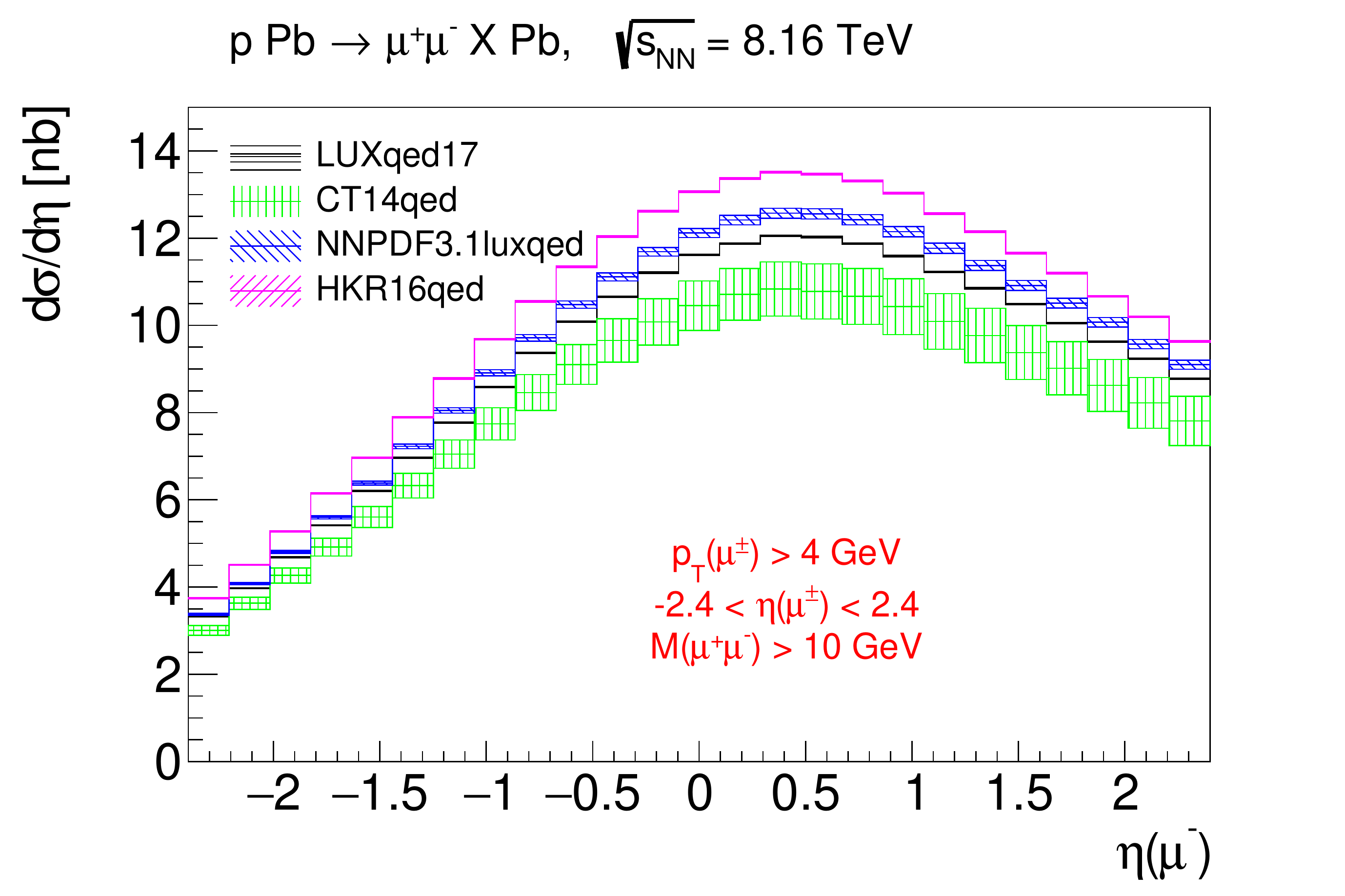}
\includegraphics[width=0.43\textwidth]{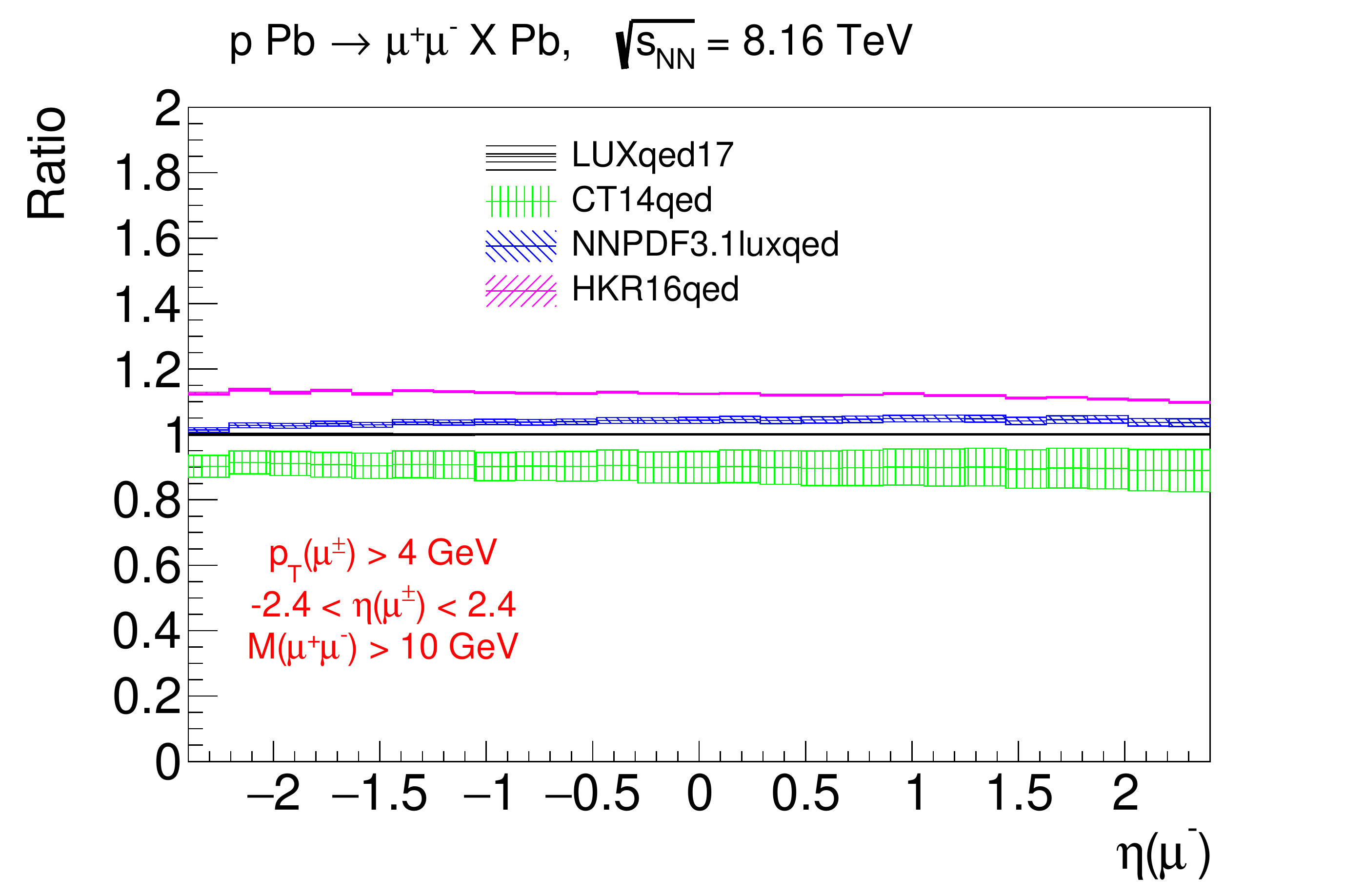}
\caption{Differential cross sections in the fiducial region for $p+\textrm{Pb}\rightarrow \textrm{Pb} + \ell^+\ell^- + X$ production at $\sqrt{s_{N N}} = 8.16$~\TeV\ for different collinear photon PDF sets.
Four differential distributions are shown (from top to bottom): invariant mass of lepton pair, pair rapidity, transverse momentum of negatively-charged lepton and its pseudorapidity. Figures on the right show the ratios to LUXqed17 PDF. The bands denote the PDF uncertainties (if available) calculated at 68\% CL, and the statistical uncertainties of the calculations added in quadrature.}
\label{fig:inc_cut}
\end{figure}

The integrated fiducial cross-sections for $p+\textrm{Pb}\rightarrow \textrm{Pb} + \ell^+\ell^- + X$ production at $\sqrt{s_{N N}} = 8.16$~\TeV\ for different collinear photon PDF sets are summarized in Tab.~\ref{fig:xs}.
Comparison of several lepton kinematic distributions between different photon-PDFs is shown in Fig.~\ref{fig:inc_cut}, including invariant mass and rapidity of lepton pair, and single-lepton transverse momentum/pseudorapidity distributions. 
The asymmetry visible in pair rapidity and single-lepton pseudorapidity distributions is due to expermental setup, which assumes a difference in the energy per nucleon between the proton beam and the lead beam (see Sec.~\ref{sec:experiment}).
All photon PDF parameterizations agree within 20\% with each other.
The differences are mainly due to overall PDF normalization, as no variation in the shape of various kinematic distributions is observed.

To check the sensitivity to the nuclear form factor modelling (Eq.~\ref{pb_el_flux}), different values of $R_A$ ($R_A = 7.1$ fm) and $a$ ($a=0.55$ fm) parameters are used, in a similar way as in Ref.~\cite{Azevedo:2019fyz}. These variations change the fiducial cross-sections by 4\% and 3\% respectively.

\begin{table}[h!]
\begin{center}
\begin{tabular}{|l|c|c|}
\hline
Contribution & $p_T^{\ell} > 4$ \GeV & $p_T^{\ell}  > 4$ \GeV, $|\eta^{\ell}| < 2.4$,\\
& & $m_{\ell^+\ell^-} > 10$ \GeV\\
\hline
$\gamma^{p}_{\rm{el}}$  & 44.9 nb & 17.5 nb\\ 
\hline
$\gamma^{p}_{\rm{el}} + \gamma^{p}_{\rm{inel}}$ [CT14qed\_inc] & $98\pm4$ (PDF) nb & $40\pm 2$ (PDF) nb\\
\hline
$\gamma^{p}_{\rm{el}} + \gamma^{p}_{\rm{inel}}$ [LUXqed17] & $105.8\pm0.2$ (PDF) nb & $44.1 \pm 0.1$ (PDF) nb\\
\hline
$\gamma^{p}_{\rm{el}} + \gamma^{p}_{\rm{inel}}$ [NNPDF3.1luxQED] & $115.6 \pm 0.6$ (PDF) nb & $45.9 \pm 0.3$ (PDF) nb\\
\hline
$\gamma^{p}_{\rm{el}} + \gamma^{p}_{\rm{inel}}$ [HKR16qed] & 121.6 nb & 49.4 nb\\
\hline
\end{tabular}
\end{center}
\caption{Integrated fiducial cross sections for $p+\textrm{Pb}\rightarrow \textrm{Pb} + \ell^+\ell^- + X$ production at $\sqrt{s_{N N}} = 8.16$~\TeV\ for different collinear photon PDF sets. 
The effect of applying only $p_T^{\ell}$ requirement is shown in second column.
The uncertainties denote the PDF uncertainties (if available) calculated at 68\% CL.
For comparison, the cross section for purely elastic contribution is also shown.}
\label{fig:xs}
\end{table}

\section{Results using $k_T$-factorization approach}

Several different parametrizations of proton strucure functions are used. Those are labeled as:
\begin{itemize} 

 \item ALLM \cite{Abramowicz:1991xz,Abramowicz:1997ms}: This parametrization gives a good fit to $F_2$ in most of the measured regions.

\item SY \cite{Suri:1971yx}: This parameterization of Suri and Yennie from the early 1970's does not include DGLAP evolution. It is still  used as one of the defaults in the LPAIR event generator~\cite{Vermaseren:1982cz}.
    
\item SU \cite{Szczurek:1999rd}: A parametrization which concentrates to give a good description at small and intermediate $Q^2$ for $x > 0.01$.
At large $Q^2$, it is complemented by the NNLO calculation of $F_2$ and $F_L$ from NNLO MSTW 2008 PDF analysis~\cite{Martin:2009iq}.
   
\item LUX-like: a recently constructed parametrization, described in details in Ref.~\cite{Luszczak:2018ntp}.
This setup closely follows the LUXqed work from Ref.\cite{Manohar:2017eqh}.
\end{itemize} 

To model $\gamma^{p}_{el}(x, Q^2)$ we use Eq.~\ref{proton_el_flux} with so-called dipole parametrization of the proton form factors:
\begin{eqnarray}
G_E(Q^2) &=& \left( 1+\frac{Q^2}{Q_0^2} \right)^{-2} \\
G_M(Q^2) &=& \mu_p G_E(Q^2)~,
\end{eqnarray}
where $\mu_p$ is the proton magnetic moment.

Table~\ref{tab:kt} shows the comparison of integrated fiducial cross sections for inelastic $p+\textrm{Pb}\rightarrow \textrm{Pb} + \ell^+\ell^- + X$ production at $\sqrt{s_{N N}} = 8.16$~\TeV\ for different proton structure functions.
All structure functions provide similar fiducial cross-section, at the level of 16--18 nb.
These inelastic cross-sections are also similar in size to the elastic contribution (18 nb) and are slightly lower than the numbers from collinear analysis, subtracted for elastic part (see Table~\ref{fig:xs}).
A comparison is also made with LUX-like parametrization when the longitudinal structure function ($F_L$) is explicitly considered.
This leads to the decrease of the cross section by 2\%, similarly to Ref.~\cite{Luszczak:2018ntp}.

\begin{figure}[!h]
 \includegraphics[width=0.49\textwidth]{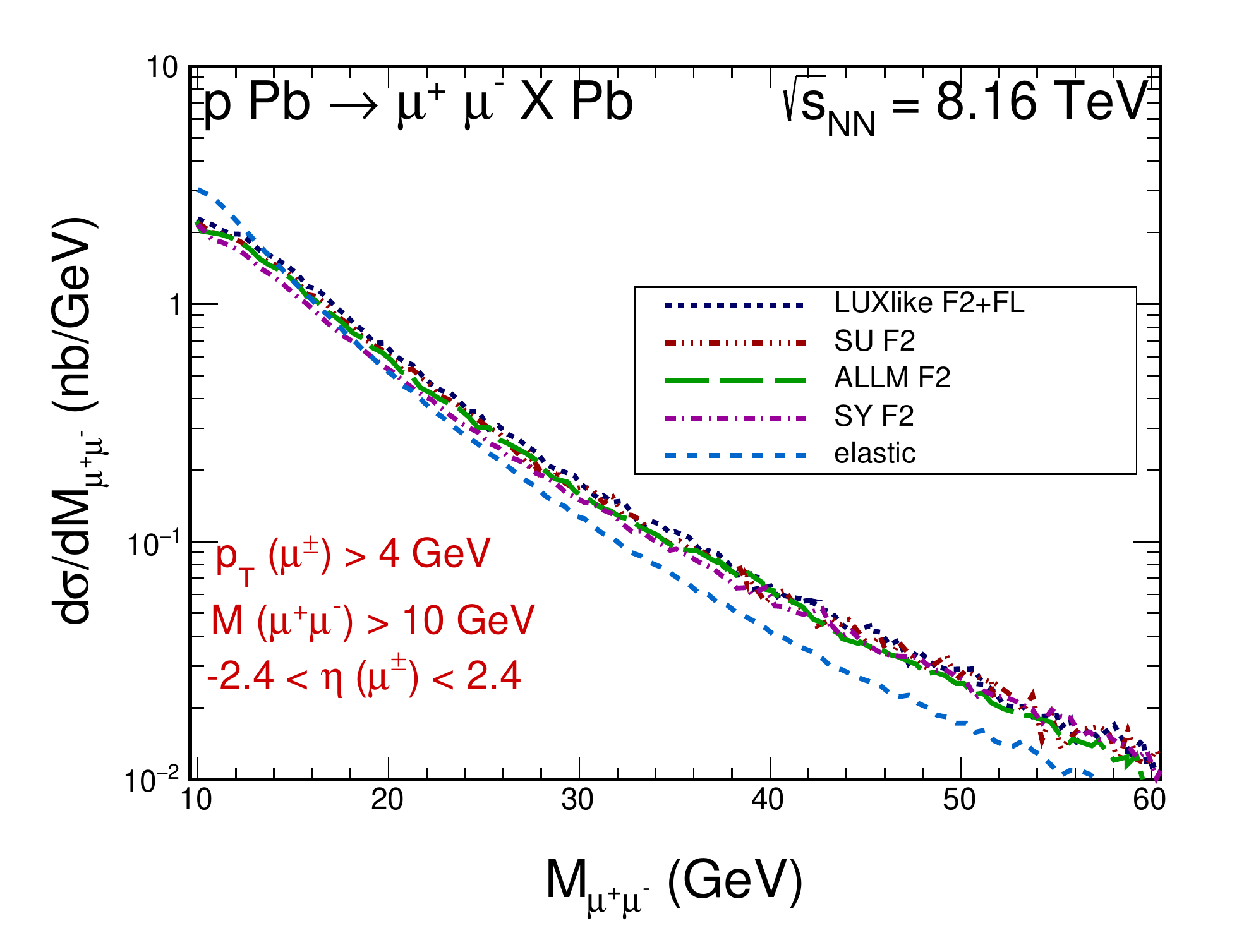}
  \includegraphics[width=0.49\textwidth]{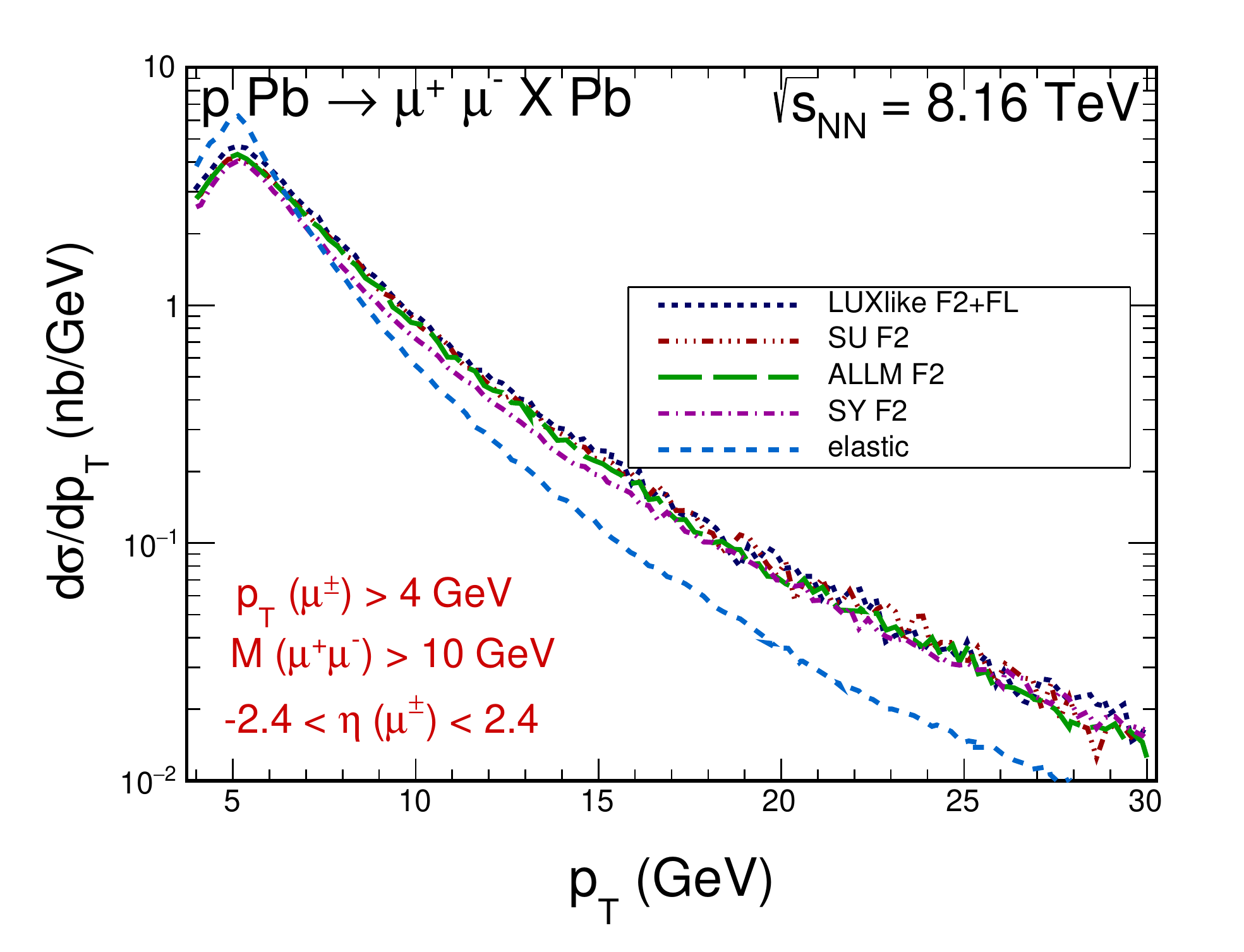}
 \includegraphics[width=0.49\textwidth]{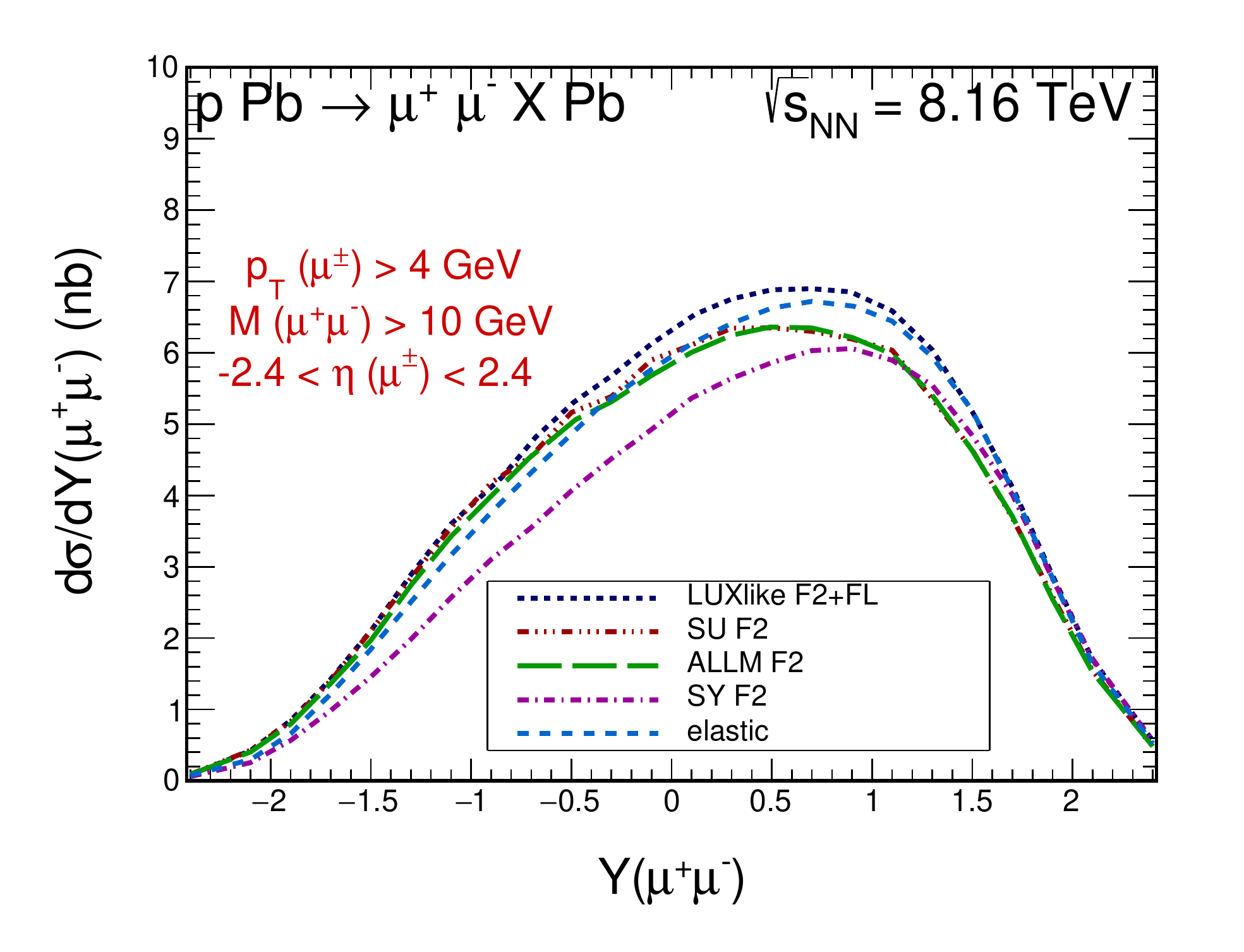}
  \includegraphics[width=0.49\textwidth]{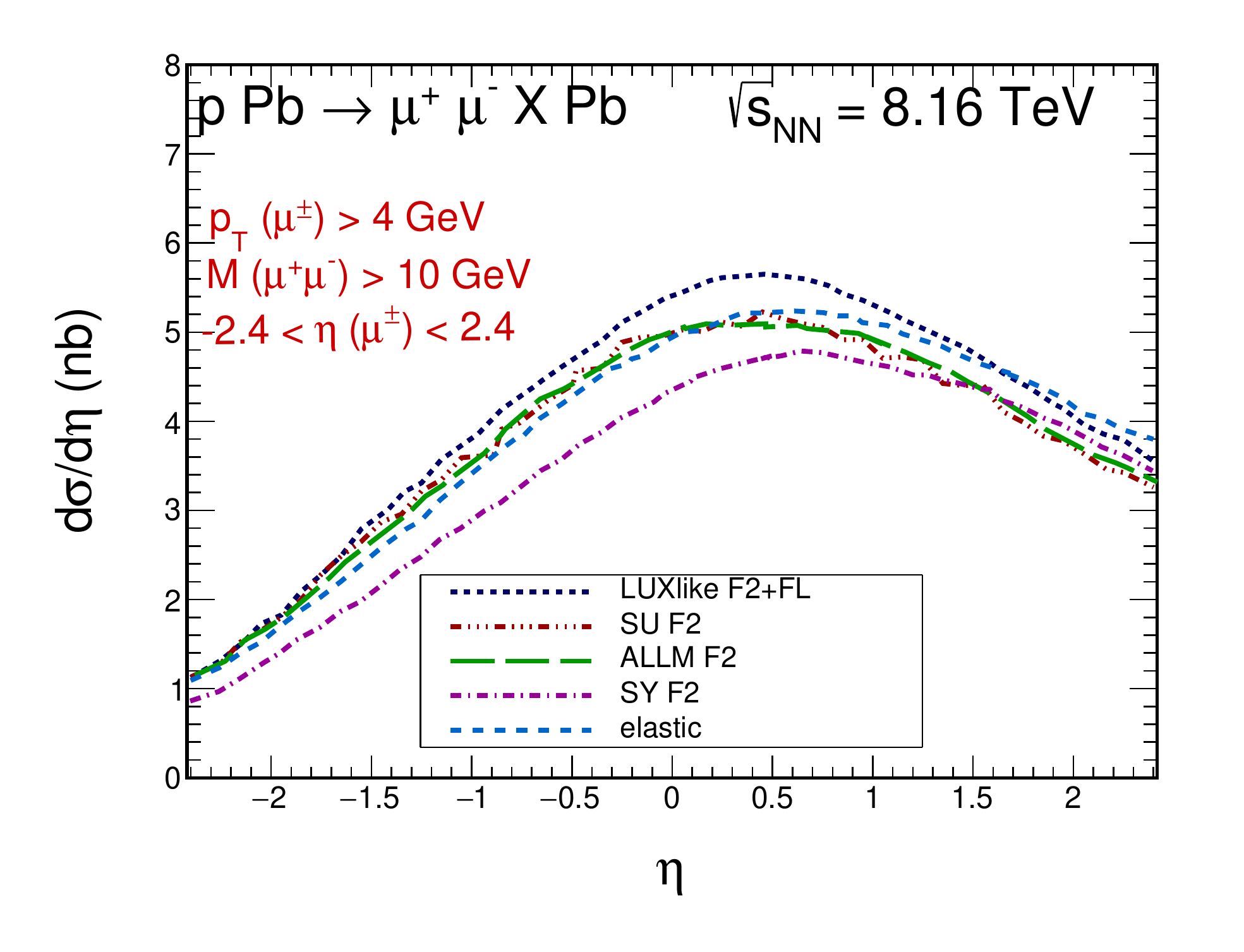}
\caption{Differential cross sections in the fiducial region for $p+\textrm{Pb}\rightarrow \textrm{Pb} + \ell^+\ell^- + X$ production at $\sqrt{s_{N N}} = 8.16$~\TeV\ in $k_T$ factorization approach for several proton structure functions.
Four differential distributions are shown: invariant mass of lepton pair (top left), leading lepton transverse momentum (top right),
dilepton rapidity (bottom left) and leading lepton pseudorapidity (bottom right).
For comparison, the elastic contribution ($p+\textrm{Pb}\rightarrow p+ \textrm{Pb} + \ell^+\ell^-$) is also shown.
}
 \label{fig:kt_figures1}
\end{figure}

\begin{figure}[!h]
\includegraphics[width=0.49\textwidth]{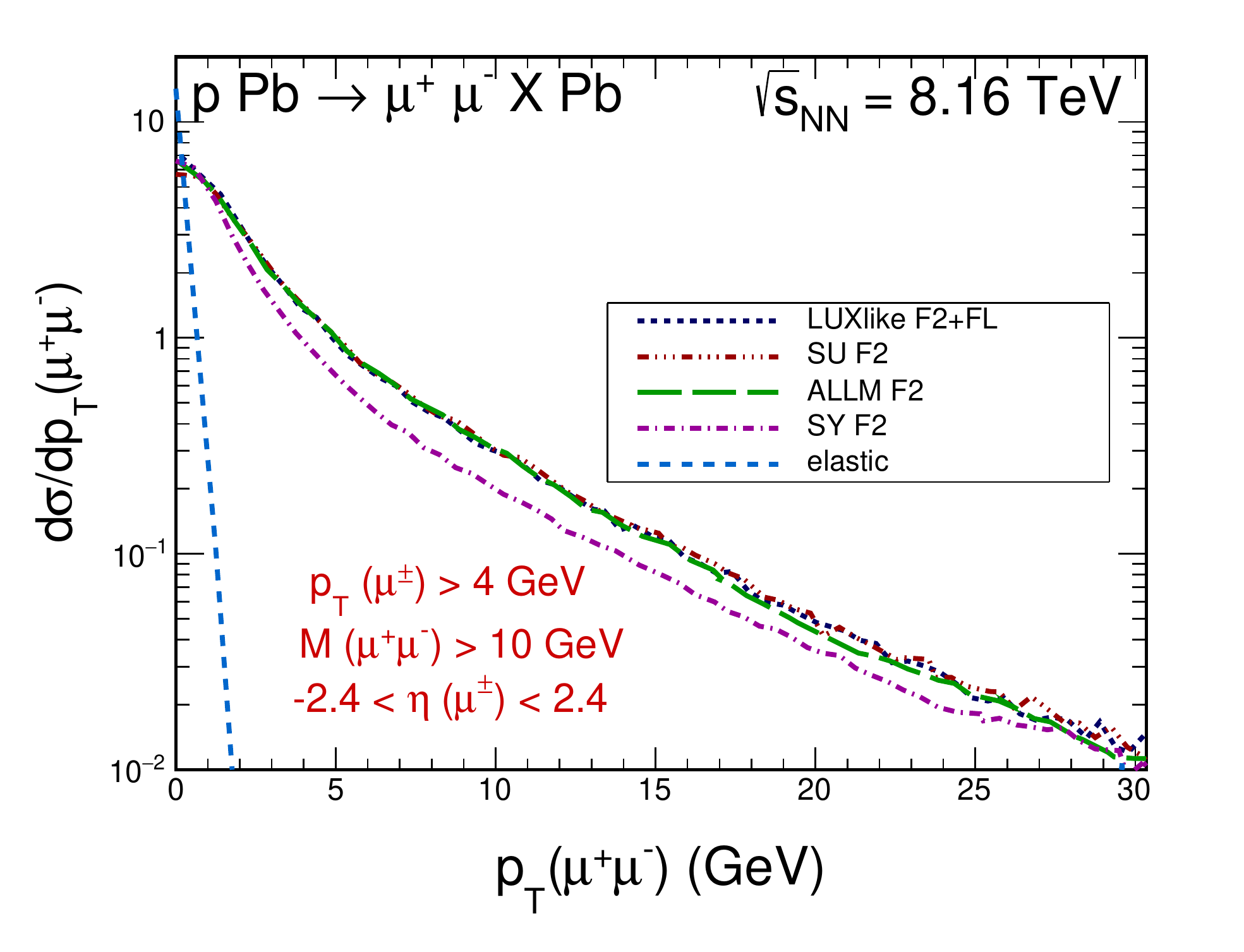}
 \includegraphics[width=0.49\textwidth]{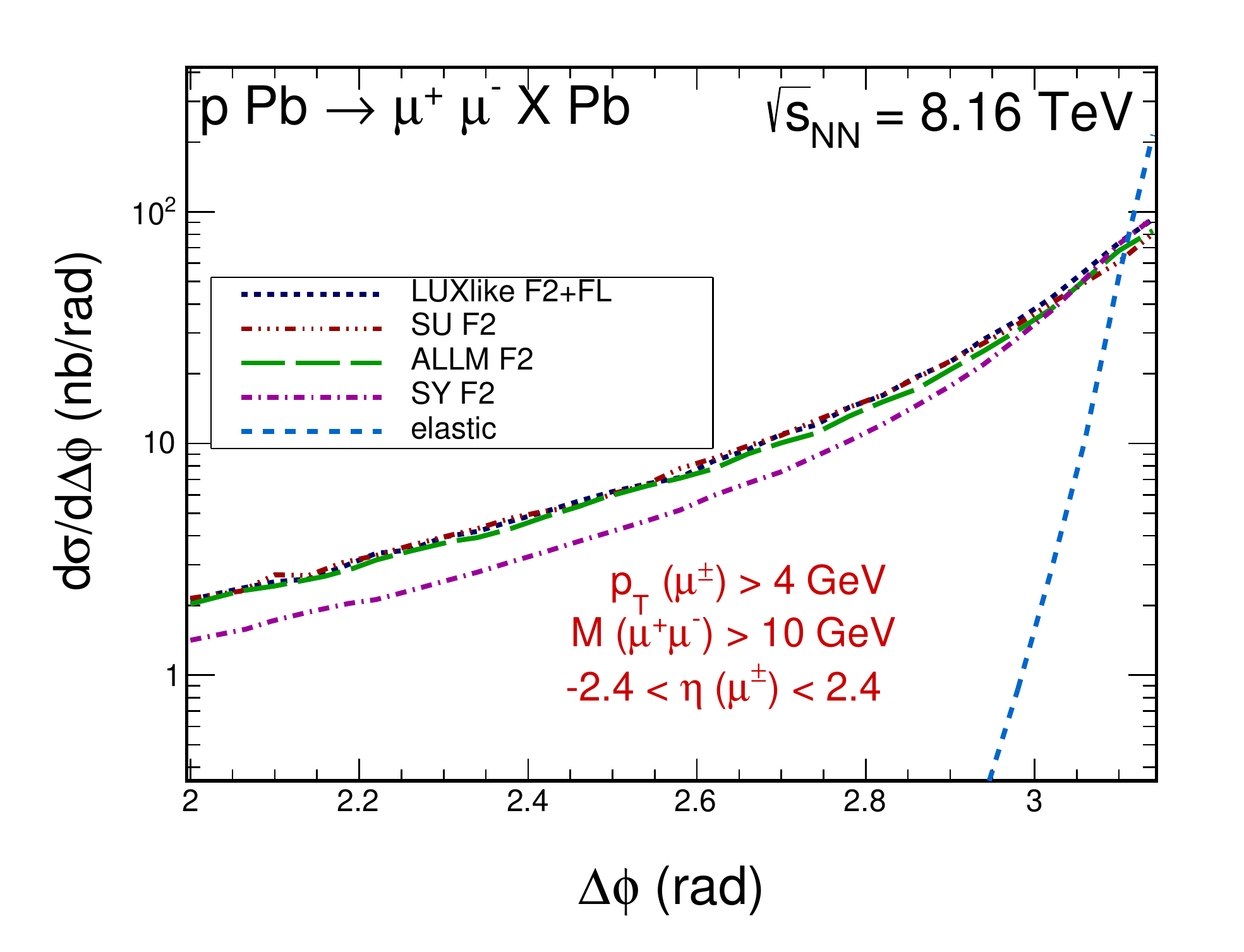}
\caption{Differential cross sections in the fiducial region for $p+\textrm{Pb}\rightarrow \textrm{Pb} + \ell^+\ell^- + X$ production at $\sqrt{s_{N N}} = 8.16$~\TeV\ in $k_T$ factorization approach for several proton structure functions.
Two differential distributions are shown: transverse momentum of lepton pair (left) and azimuthal angle difference between the pair (right).
For comparison, the elastic contribution ($p+\textrm{Pb}\rightarrow p+ \textrm{Pb} + \ell^+\ell^-$) is also shown.
}
 \label{fig:kt_figures2}
\end{figure}

Figure~\ref{fig:kt_figures1} presents differential cross sections for several lepton kinematic distributions: invariant mass of lepton pair, leading lepton transverse momentum, lepton pseudorapidity difference and leading lepton pseudorapidity.
The shapes of the distributions obtained with various proton structure functions are very similar.
For completeness, differential cross sections as a function of lepton pair transverse momentum and azimuthal angle difference between the pair  are shown in Fig.~\ref{fig:kt_figures2}. Quite large (small) transverse momenta (angle differences) are possible, in contrast to leading-order calculations with collinear photons where the corresponding distributions are just Dirac delta functions. 
The $k_T$-factorization approach should be considered more appropriate here. It is also visible that the SY parametrization gives lower predictions at larger pair-$p_T$, comparing to the other parametrizations used. This is because SY parametrization does not include explicit DGLAP evolution terms, which are relevant for large photon virtualities.

\begin{figure}[!h]

 \includegraphics[width=0.49\textwidth]{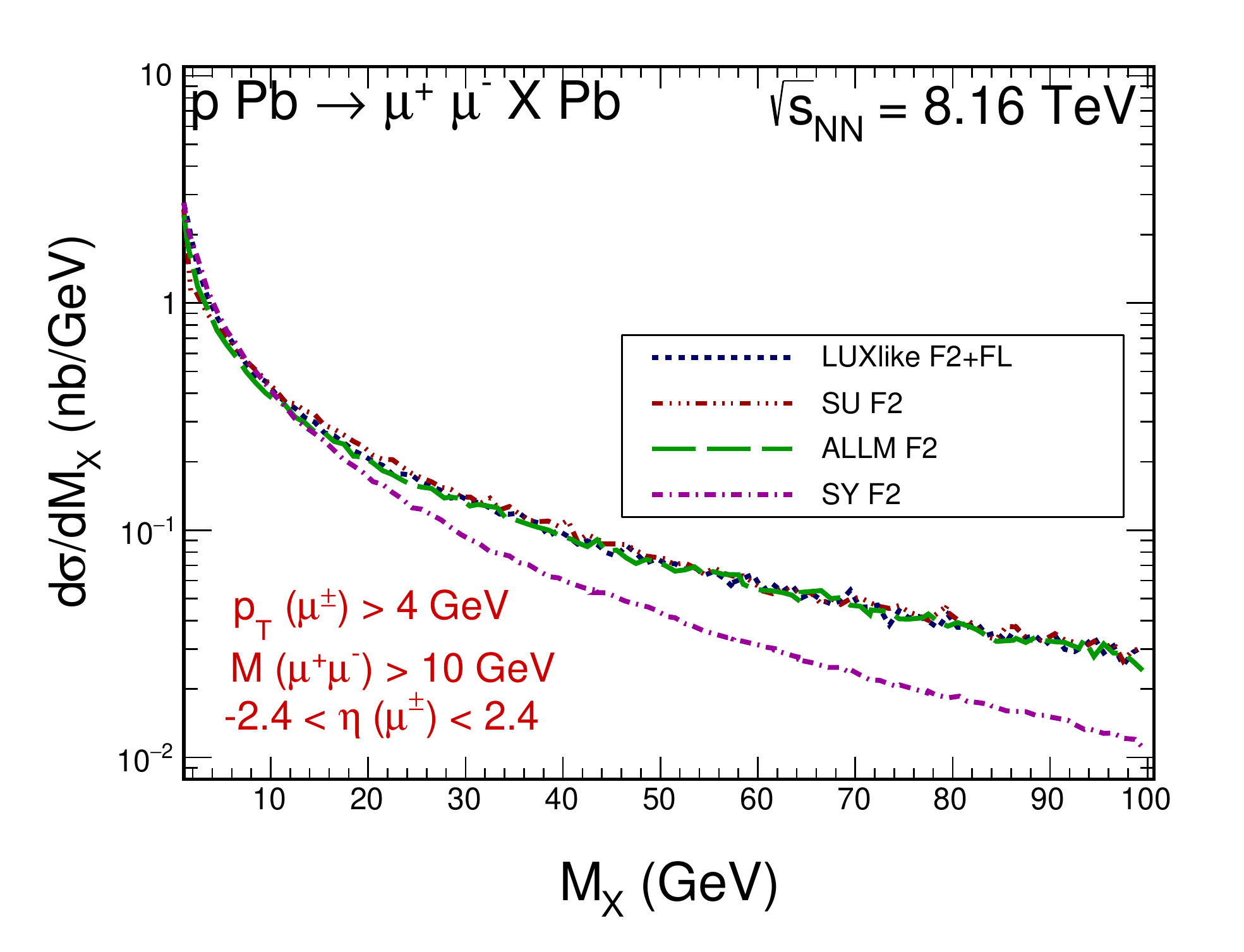}
\caption{Differential cross section as a function of the mass of the proton remnants in the fiducial region for $p+\textrm{Pb}\rightarrow \textrm{Pb} + \ell^+\ell^- + X$ production at $\sqrt{s_{N N}} = 8.16$~\TeV\ in $k_T$ factorization approach for several proton structure functions.
}
 \label{fig:kt_figures3}
\end{figure}

Based on Fig.~\ref{fig:kt_figures2}, it is also possible to separate experimentally the elastic part ($p+\textrm{Pb}\rightarrow p+ \textrm{Pb} + \ell^+\ell^-$), with striking back-to-back topology, from the inelastic contribution.
With $k_T$-factorization, one can also calculate the mass of the proton remnants ($M_X$). This is shown in Fig.~\ref{fig:kt_figures3}; in contrast to the elastic case ($M_X=m_p$) quite large masses of the remnant system can be achieved.

\begin{table}[t]
\centering
\begin{tabular}{|l|c|c|c|}
\hline
Contribution  &  $p_{T}^{\ell}>4~\GeV$ & $p_{T}^{\ell}>4~\GeV$, $|\eta^{\ell}|<2.4 $, $m_{\ell^+\ell^-}>10~\GeV$ \\
\hline
$\gamma^{p}_{\rm{el}}$   & 47.9 nb  & 18.3 nb \\
\hline
$\gamma^{p}_{\rm{inel}}$ [LUX-like  $F_2$]  & 43.6 nb  &  17.4 nb\\
\hline
$\gamma^{p}_{\rm{inel}}$ [LUX-like  $F_2+F_L$]  & 42.6 nb    & 17.1 nb\\
\hline    
$\gamma^{p}_{\rm{inel}}$ [ALLM97 $F_2$]  & 41.7 nb   &16.4 nb\\
\hline
$\gamma^{p}_{\rm{inel}}$ [SU $F_2$]  & 41.7  nb &16.7 nb\\
\hline 
$\gamma^{p}_{\rm{inel}}$ [SY $F_2$]  & 40.4  nb  &16.0 nb\\
\hline
\end{tabular}
\caption{Integrated fiducial cross sections for inelastic $p+\textrm{Pb}\rightarrow \textrm{Pb} + \ell^+\ell^- + X$ production at $\sqrt{s_{N N}} = 8.16$~\TeV\ for different proton structure functions. 
The effect of applying only $p_T^{\ell}$ requirement is shown in second column.
}
\label{tab:kt}
\end{table}

\clearpage
\section{Discussion}
\label{sec:discussion}

\begin{figure}[]
\includegraphics[width=0.43\textwidth]{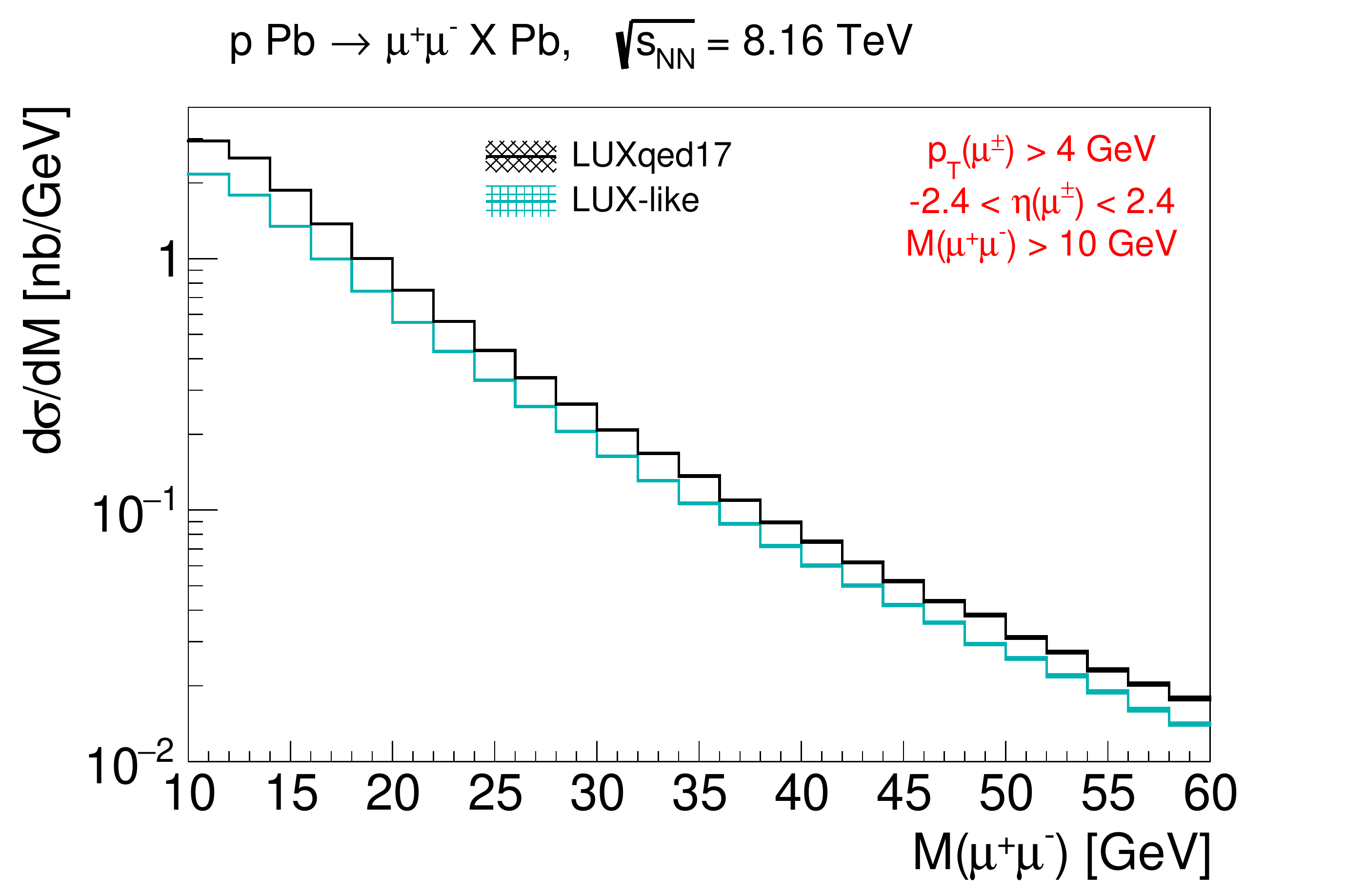}
\includegraphics[width=0.43\textwidth]{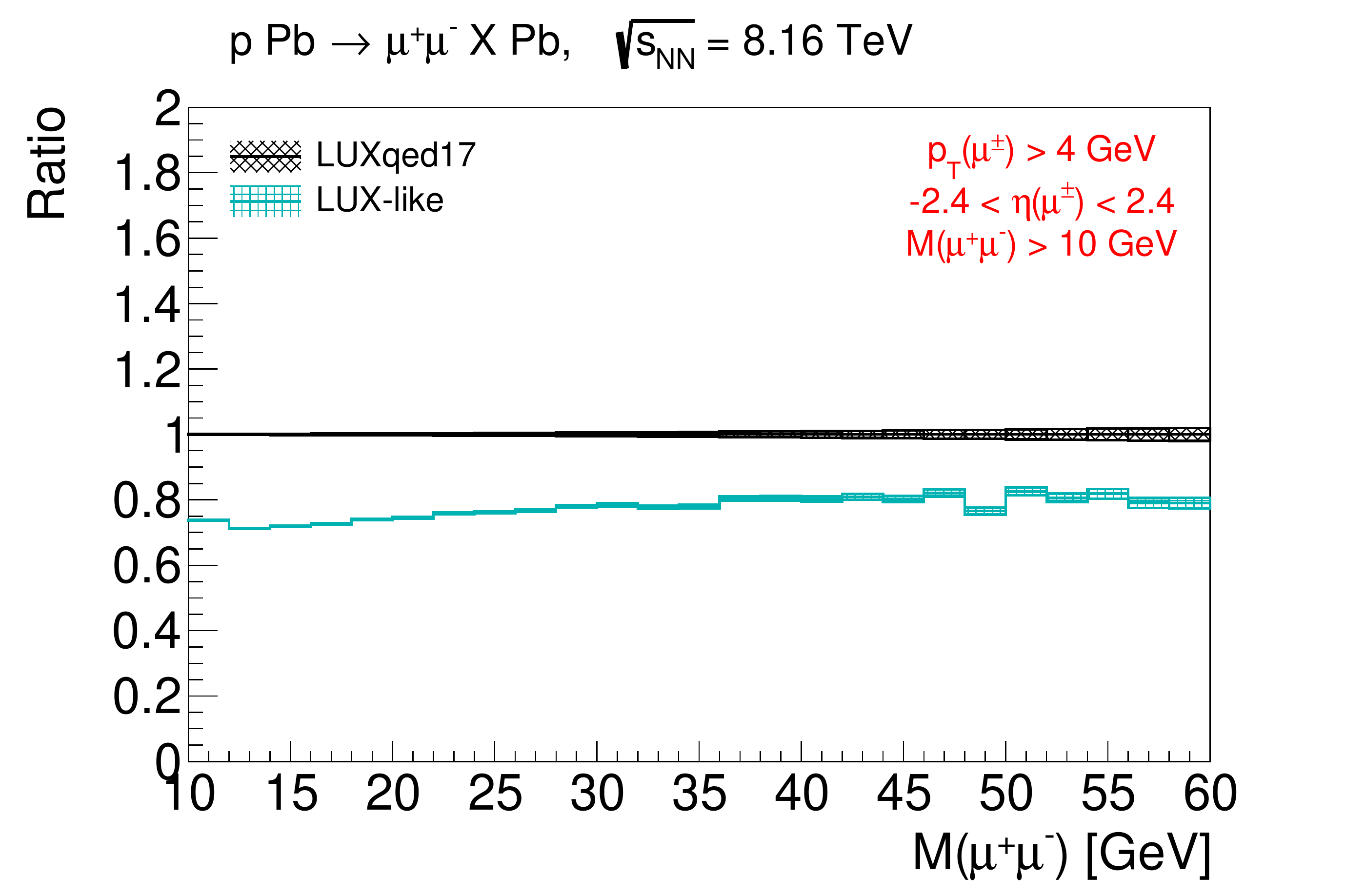}
\includegraphics[width=0.43\textwidth]{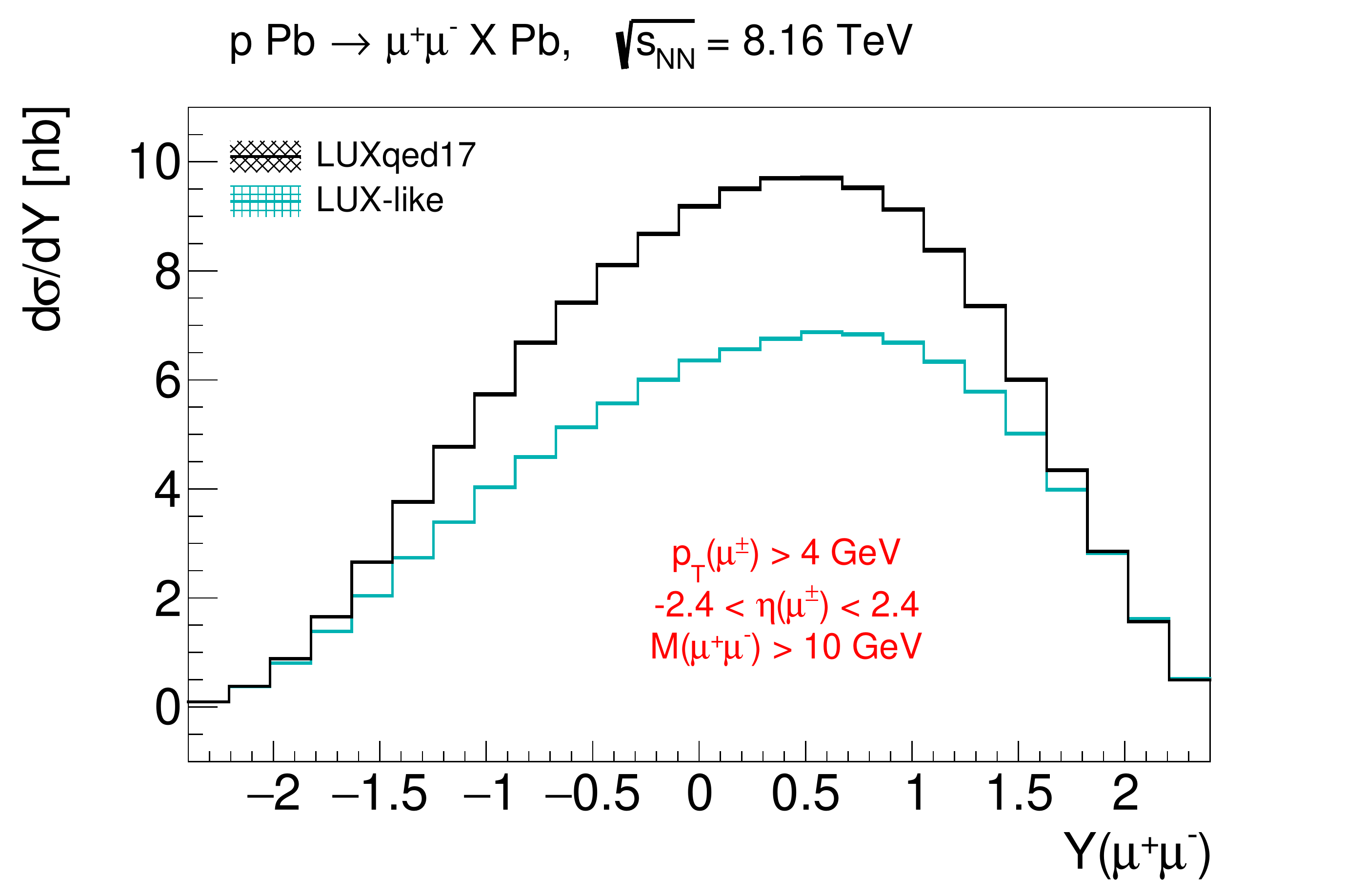}
\includegraphics[width=0.43\textwidth]{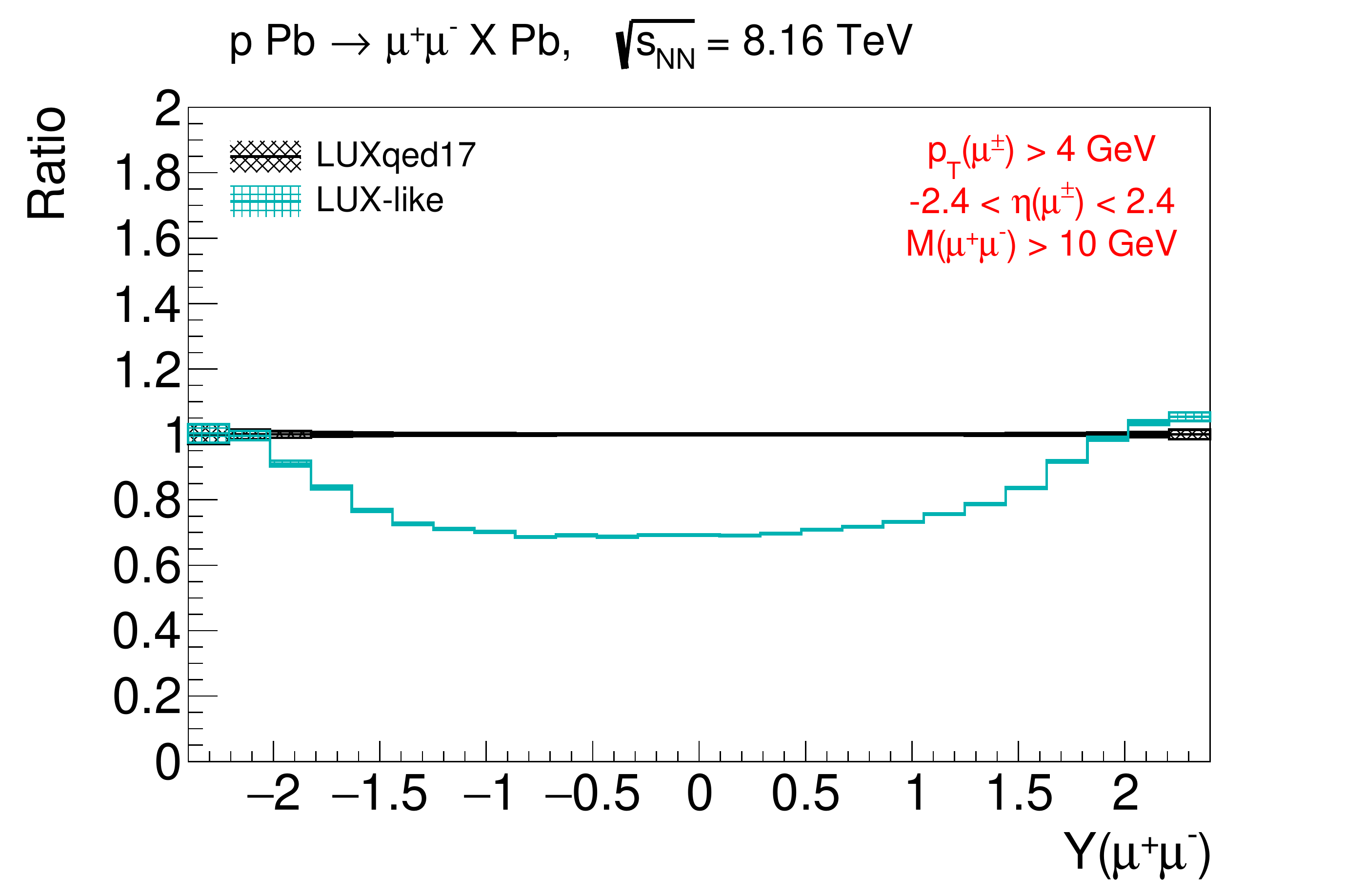}
\includegraphics[width=0.43\textwidth]{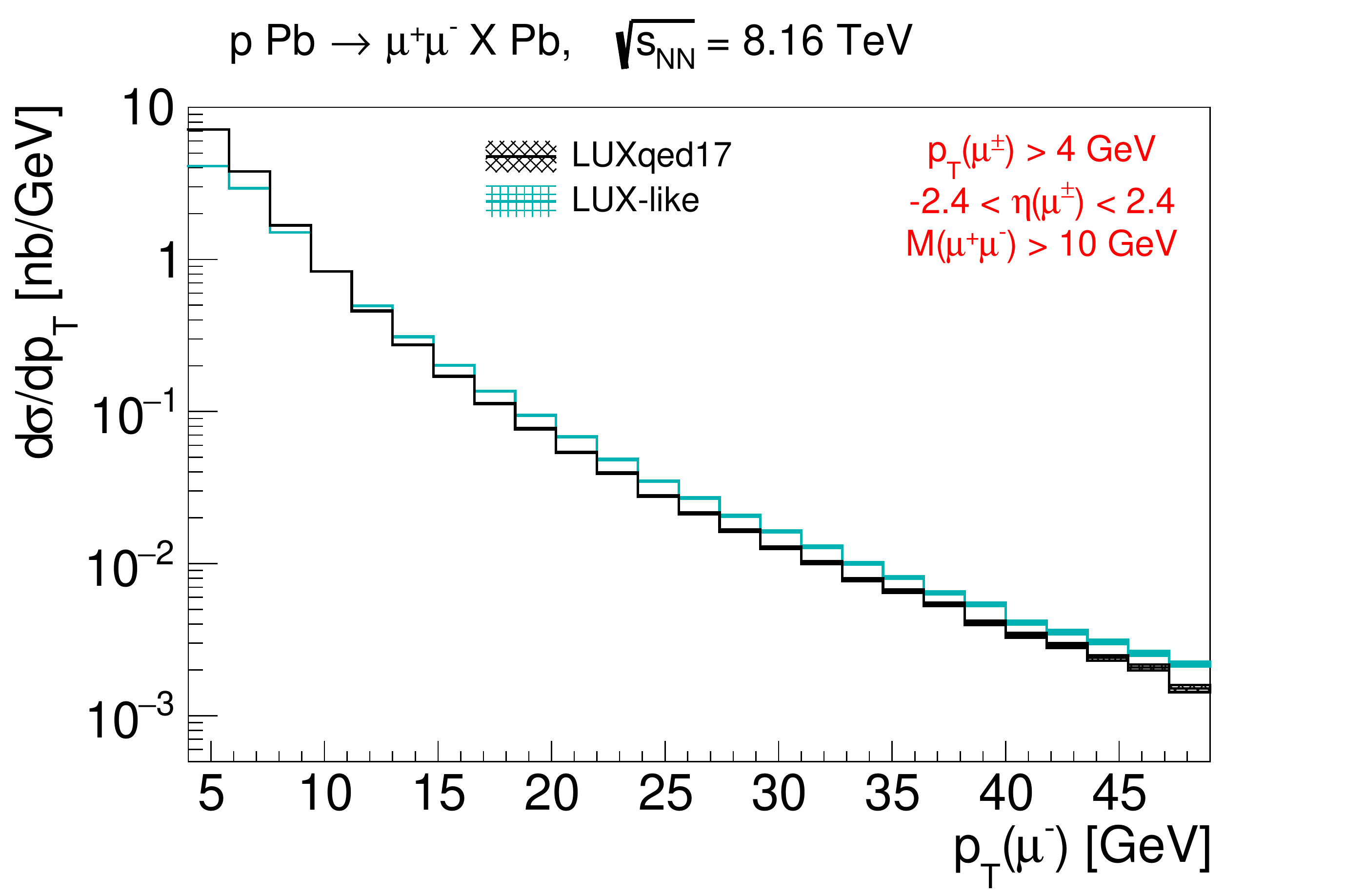}
\includegraphics[width=0.43\textwidth]{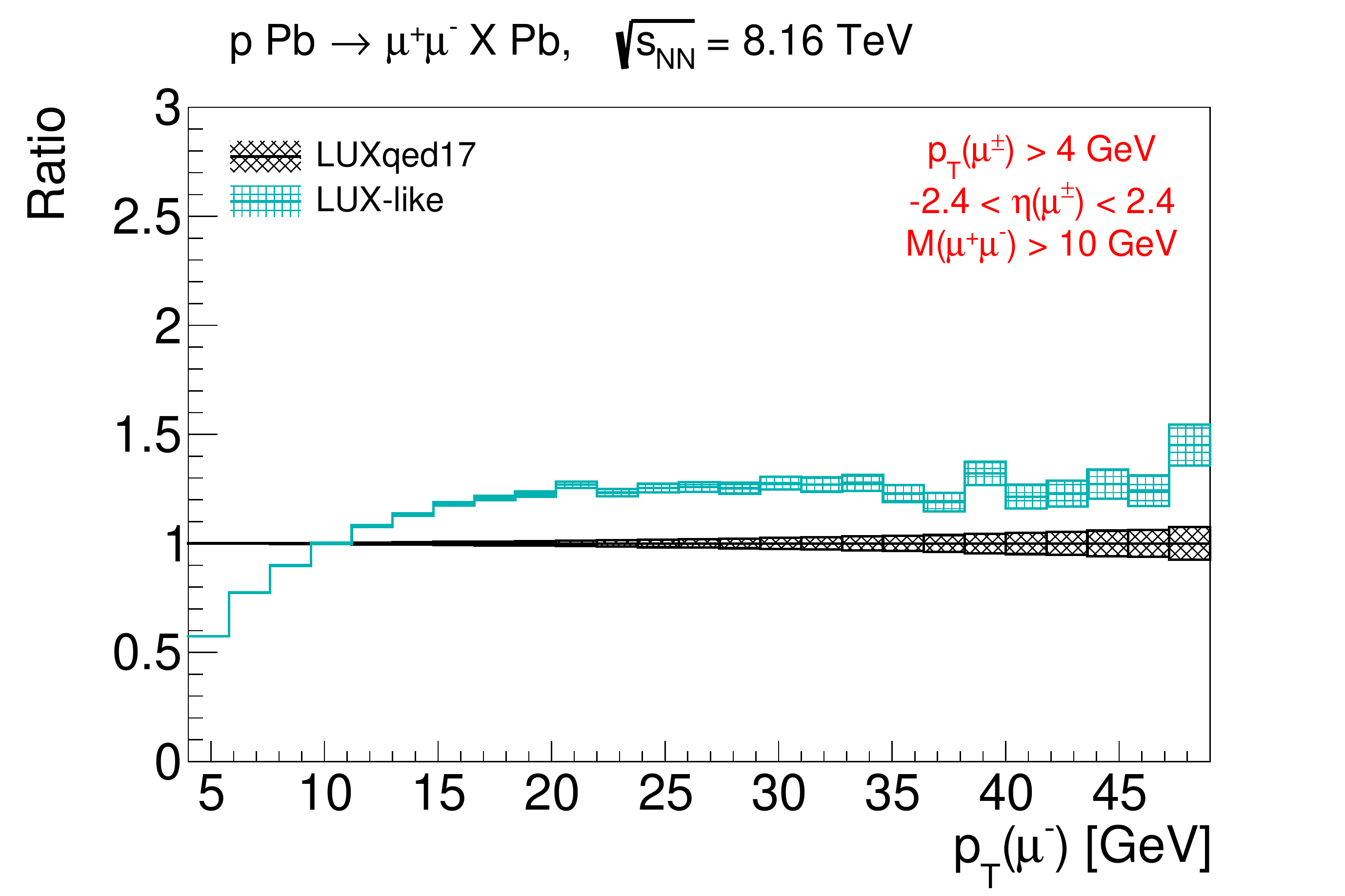}
\includegraphics[width=0.43\textwidth]{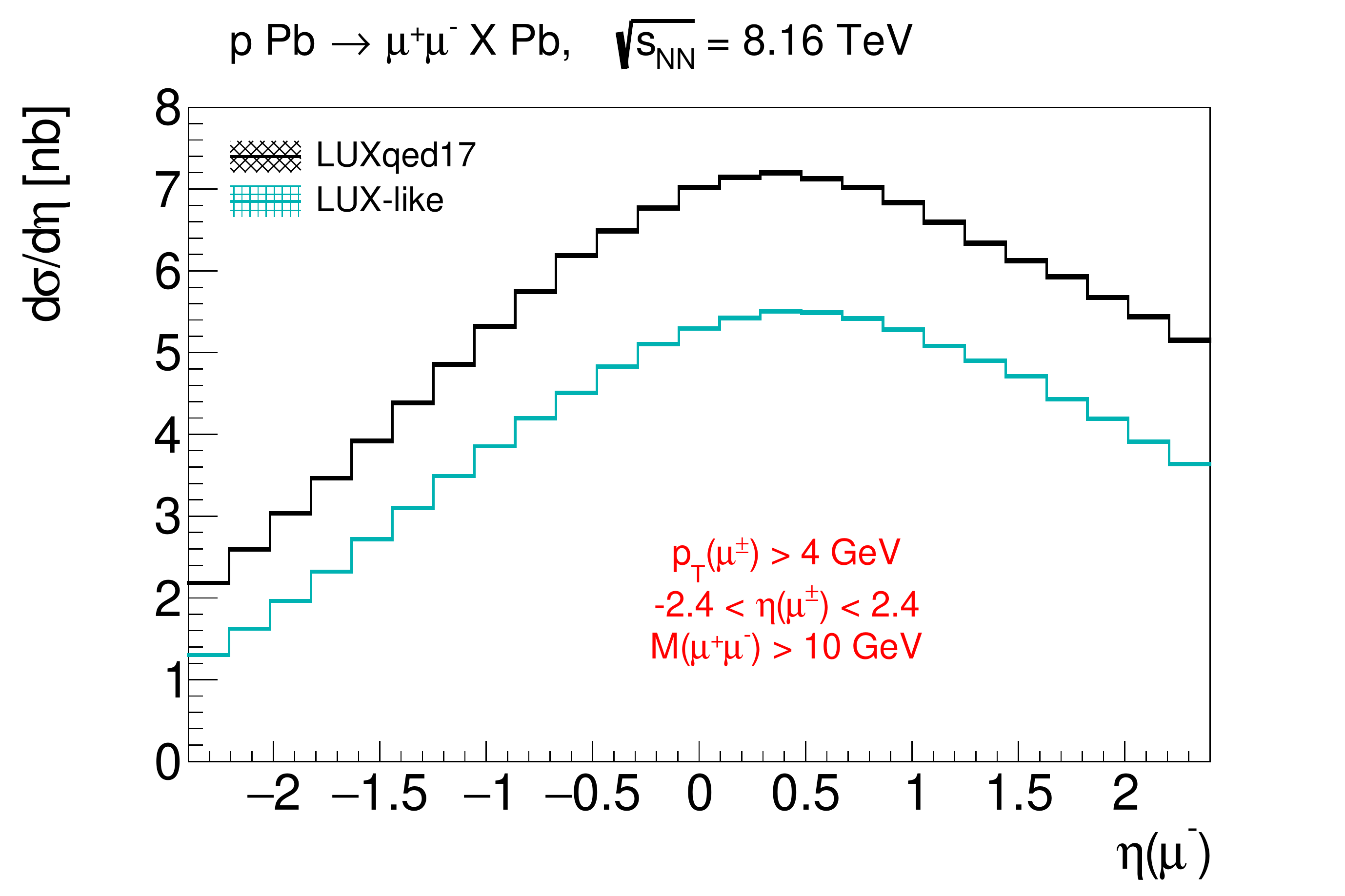}
\includegraphics[width=0.43\textwidth]{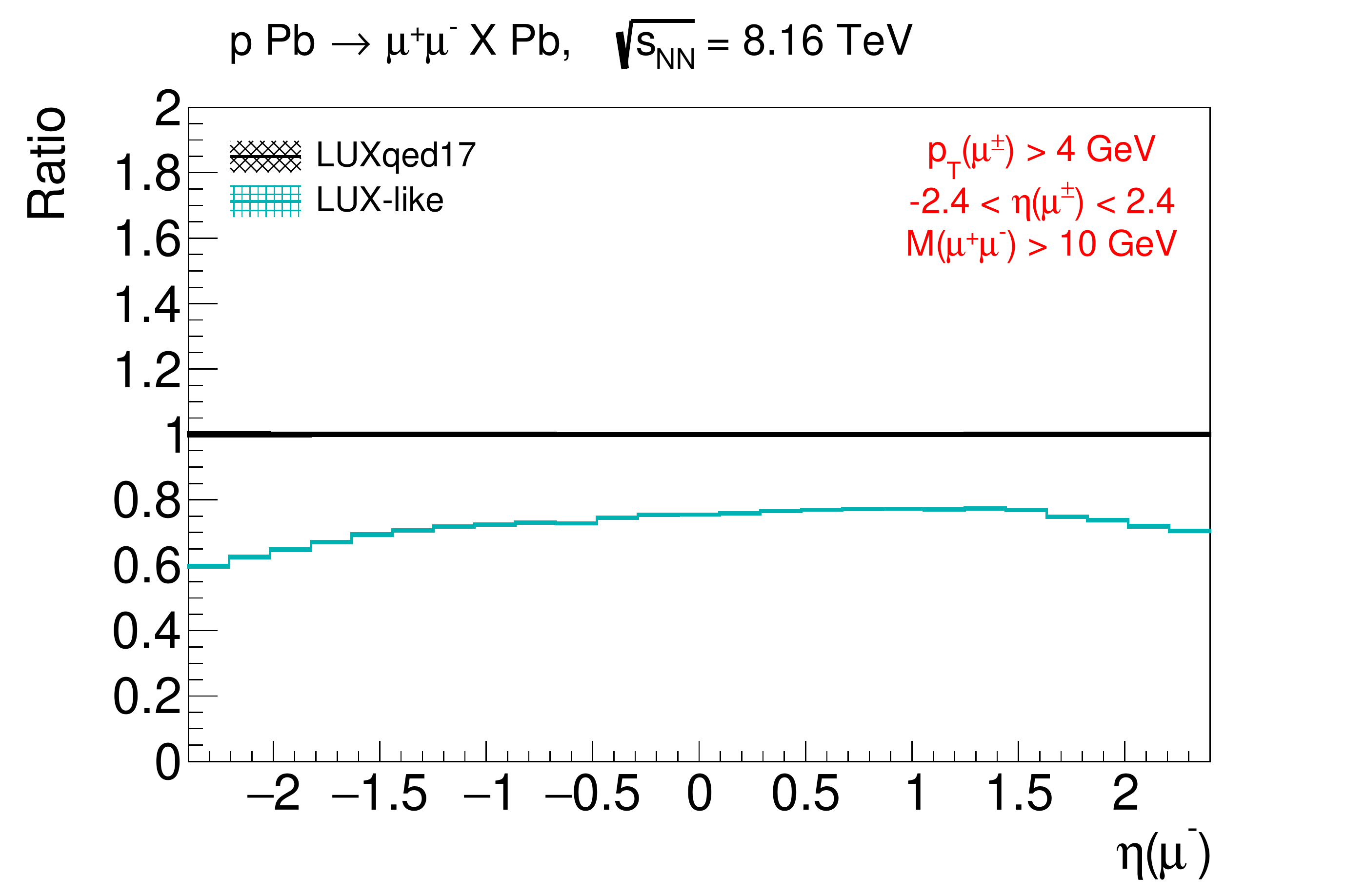}
\caption{Differential cross sections in the fiducial region for $p+\textrm{Pb}\rightarrow \textrm{Pb} + \ell^+\ell^- + X$ production at $\sqrt{s_{N N}} = 8.16$~\TeV\ for collinear LUXqed17 photon PDF and
for LUX-like $F_2+F_L$ photon PDF with $k_T$-factorization.
Four differential distributions are shown (from top to bottom): invariant mass of lepton pair, pair rapidity, transverse momentum of negatively-charged lepton and its pseudorapidity. Figures on the right show the ratios to LUXqed17 PDF.}
\label{fig:inc_cut_kt}
\end{figure}

Figure~\ref{fig:inc_cut_kt} compares several differential distributions computed using the two approaches. For the collinear approach pure inelastic contribution is estimated by
subtracting elastic part computed following Eq.~\ref{eq:elasticRenat}.
For the invariant mass distribution and lepton pseudorapidity the shapes are similar and
the main difference between the two predictions is observed in the normalization.
For the distribution of the lepton pair rapidity the two predictions agree at larger rapidities while disagreement concentrates
in the central region. The biggest difference is observed for the transverse momentum distribution of the lepton where at low $p_T$ collinear approximation exceeds the estimate
from $k_T$-factorization approach while at high $p_T$ the ordering is reversed.  
This suggests that at low $p_T$ (close to the boundary of the fiducial region) the difference is due to the smearing of dilepton transverse momentum introduced by the $k_T$-factorization approach.

We also take the opportunity to calculate expected number of events for realistic assumption on total integrated luminosity.
Based on the previous $p+\textrm{Pb}$ runs at the LHC, we assume  $\int Ldt= 200~\textrm{nb}^{-1}$.
We also assume possible experimental efficiencies, mainly due to trigger and reconstruction of leptons, which we embed in a single correction factor $C=0.7$.

Table~\ref{fig:numbers} shows the expected number of events for $p+\textrm{Pb}\rightarrow \textrm{Pb} + \ell^+\ell^- + X$ production at $\sqrt{s_{N N}} = 8.16$~\TeV\ and configuration described above. 
Approximately 2500 elastic dilepton events are expected. 
Depending on the calculations, 3400 (collinear with LUXqed17 PDF) or 2400 ($k_T$-factorization with LUX-like $F_2+F_L$) reconstructed inelastic events are predicted. 
The data should be therefore sensitive to discriminate between the predictions based on  
collinear and $k_T$-factorization approaches, using existing datasets collected by ATLAS and CMS.

\begin{table}[t]
\begin{center}
\begin{tabular}{|l|c|c|}
\hline
Contribution & Expected events ($C=1$) & Expected events ($C=0.7$) \\
\hline
$\gamma^{p}_{\rm{el}}$  & 3600 & 2500\\ 
\hline
$\gamma^{p}_{\rm{inel}}$ [LUXqed17 collinear] & 5600 & 3900 \\
\hline
$\gamma^{p}_{\rm{inel}}$ [LUX-like $F_2+F_L$] & 3400 & 2400 \\
\hline
\end{tabular}
\end{center}
\caption{Expected number of events for $p+\textrm{Pb}\rightarrow \textrm{Pb} + \ell^+\ell^- + X$ production at $\sqrt{s_{N N}} = 8.16$~\TeV\ assuming $\int Ldt= 200~\textrm{nb}^{-1}$. 
Shown are several contributions: purely elastic, inelastic with collinear LUXqed17 PDF and inelastic with $k_T$-factorization and LUX-like $F_2+F_L$ proton structure function parameterization.
An effect of possible experimental efficiencies is shown in last column.}
\label{fig:numbers}
\end{table}

\FloatBarrier
\section{Summary}

In summary, we propose a method that would provide an unambiguous test of the photon parton distribution at LHC energies, and allow 
constraints to be placed on it.
This method is based on the measurement of the cross-section for the reaction $p+\textrm{Pb}\rightarrow \textrm{Pb} + \ell^+\ell^- + X$, where the expected background is small compared to the analogous process in $pp$ collisions. 
Results are shown for different choices of collinear photon PDFs, and a comparison is made with unintegrated photon distributions that include non-zero photon transverse momentum.
Due to the smearing of dilepton transverse momentum introduced by the $k_T$-factorization approach, these two approaches lead to the cross sections that differ by about 30\%.
Moreover, for collinear approach and  by analogy to DIS, an optimal choice of the scale is identified.
Using simple (realistic) experimental requirements on lepton kinematics, it is shown that one can expect O(3000) inelastic events with the existing datasets recorded by ATLAS/CMS at $\sqrt{s_{N N}} = 8.16$~\TeV\ for each lepton flavour.

\section*{Acknowledgements}
We would like to thank James Ferrando for useful suggestions.
The work of M.L. was partially supported by the Center for Innovation and
Transfer of Natural Sciences and Engineering Knowledge in Rzesz{\'o}w.
The work of R.S. was partially supported by the BMBF-JINR cooperation.
M.L. and R.S. acknowledge the hospitality of DESY where a portion of
this work was performed.


\section*{References}
\bibliographystyle{atlasnote}
\bibliography{main}

\providecommand{\href}[2]{#2}\begingroup\raggedright\begin{thebibliography}{10}

\bibitem{Aad:2014qja}
{ATLAS} Collaboration, G.~Aad et al., {\em {Measurement of the low-mass
  Drell-Yan differential cross section at $\sqrt{s}$ = 7 TeV using the ATLAS
  detector}\/},  \href{http://dx.doi.org/10.1007/JHEP06(2014)112}{JHEP {\bf 06}
  (2014)  112},
\href{http://arxiv.org/abs/1404.1212}{{\tt arXiv:1404.1212 [hep-ex]}}.

\bibitem{Aad:2016zzw}
{ATLAS} Collaboration, G.~Aad et al., {\em {Measurement of the
  double-differential high-mass Drell-Yan cross section in pp collisions at $
  \sqrt{s}=8 $ TeV with the ATLAS detector}\/},
  \href{http://dx.doi.org/10.1007/JHEP08(2016)009}{JHEP {\bf 08} (2016)  009},
\href{http://arxiv.org/abs/1606.01736}{{\tt arXiv:1606.01736 [hep-ex]}}.

\bibitem{Accomando:2016tah}
E.~Accomando, J.~Fiaschi, F.~Hautmann, S.~Moretti, and C.~H.
  Shepherd-Themistocleous, {\em {Photon-initiated production of a dilepton
  final state at the LHC: Cross section versus forward-backward asymmetry
  studies}\/},  \href{http://dx.doi.org/10.1103/PhysRevD.95.035014}{Phys. Rev.
  {\bf D95} (2017) no.~3, 035014},
\href{http://arxiv.org/abs/1606.06646}{{\tt arXiv:1606.06646 [hep-ph]}}.

\bibitem{Luszczak:2015aoa}
M.~Luszczak, W.~Schafer, and A.~Szczurek, {\em {Two-photon dilepton production
  in proton-proton collisions: two alternative approaches}\/},
  \href{http://dx.doi.org/10.1103/PhysRevD.93.074018}{Phys. Rev. {\bf D93}
  (2016) no.~7, 074018},
\href{http://arxiv.org/abs/1510.00294}{{\tt arXiv:1510.00294 [hep-ph]}}.

\bibitem{Harland-Lang:2016apc}
L.~A. Harland-Lang, V.~A. Khoze, and M.~G. Ryskin, {\em {The photon PDF in
  events with rapidity gaps}\/},
  \href{http://dx.doi.org/10.1140/epjc/s10052-016-4100-2}{Eur. Phys. J. {\bf
  C76} (2016) no.~5, 255},
\href{http://arxiv.org/abs/1601.03772}{{\tt arXiv:1601.03772 [hep-ph]}}.

\bibitem{Luszczak:2014mta}
M.~Luszczak, A.~Szczurek, and C.~Royon, {\em {$W^+ W^-$ pair production in
  proton-proton collisions: small missing terms}\/},
  \href{http://dx.doi.org/10.1007/JHEP02(2015)098}{JHEP {\bf 02} (2015)  098},
\href{http://arxiv.org/abs/1409.1803}{{\tt arXiv:1409.1803 [hep-ph]}}.

\bibitem{Denner:2015fca}
A.~Denner, S.~Dittmaier, M.~Hecht, and C.~Pasold, {\em {NLO QCD and electroweak
  corrections to $Z+\gamma$ production with leptonic Z-boson decays}\/},
  \href{http://dx.doi.org/10.1007/JHEP02(2016)057}{JHEP {\bf 02} (2016)  057},
\href{http://arxiv.org/abs/1510.08742}{{\tt arXiv:1510.08742 [hep-ph]}}.

\bibitem{Dyndal:2015hrp}
M.~Dyndal and L.~Schoeffel, {\em {Four-lepton production from photon-induced
  reactions in $pp$ collisions at the LHC}\/},
  \href{http://dx.doi.org/10.5506/APhysPolB.47.1645}{Acta Phys. Polon. {\bf
  B47} (2016)  1645},
\href{http://arxiv.org/abs/1511.02065}{{\tt arXiv:1511.02065 [hep-ph]}}.

\bibitem{Ababekri:2016kkj}
P.~Obul, M.~Ababekri, S.~Dulat, J.~Isaacson, C.~Schmidt, and C.~P. Yuan, {\em
  {Implications of CMS analysis of photon-photon interactions for photon
  PDFs}\/},  \href{http://dx.doi.org/10.1088/1674-1137/42/11/113101}{Chin.
  Phys. {\bf C42} (2018) no.~11, 113101},
\href{http://arxiv.org/abs/1603.04874}{{\tt arXiv:1603.04874 [hep-ph]}}.

\bibitem{Biedermann:2016guo}
B.~Biedermann, M.~Billoni, A.~Denner, S.~Dittmaier, L.~Hofer, B.~Jäger, and
  L.~Salfelder, {\em {Next-to-leading-order electroweak corrections to $pp \to
  W^+W^-\to$ 4 leptons at the LHC}\/},
  \href{http://dx.doi.org/10.1007/JHEP06(2016)065}{JHEP {\bf 06} (2016)  065},
\href{http://arxiv.org/abs/1605.03419}{{\tt arXiv:1605.03419 [hep-ph]}}.

\bibitem{Biedermann:2016yvs}
B.~Biedermann, A.~Denner, S.~Dittmaier, L.~Hofer, and B.~Jäger, {\em
  {Electroweak corrections to $pp \to \mu^+\mu^-e^+e^- + X$ at the LHC: a Higgs
  background study}\/},
  \href{http://dx.doi.org/10.1103/PhysRevLett.116.161803}{Phys. Rev. Lett. {\bf
  116} (2016) no.~16, 161803},
\href{http://arxiv.org/abs/1601.07787}{{\tt arXiv:1601.07787 [hep-ph]}}.

\bibitem{Yong:2016njr}
Y.~Wang, R.-Y. Zhang, W.-G. Ma, X.-Z. Li, and L.~Guo, {\em {QCD and electroweak
  corrections to ZZ+jet production with Z -boson leptonic decays at the
  LHC}\/},  \href{http://dx.doi.org/10.1103/PhysRevD.94.013011}{Phys. Rev. {\bf
  D94} (2016) no.~1, 013011},
\href{http://arxiv.org/abs/1604.04080}{{\tt arXiv:1604.04080 [hep-ph]}}.

\bibitem{Luszczak:2018ntp}
M.~Luszczak, W.~Schafer, and A.~Szczurek, {\em {Production of $W^+ W^-$ pairs
  via $\gamma^*\gamma^* \to W^+ W^-$ subprocess with photon transverse
  momenta}\/},  \href{http://dx.doi.org/10.1007/JHEP05(2018)064}{JHEP {\bf 05}
  (2018)  064},
\href{http://arxiv.org/abs/1802.03244}{{\tt arXiv:1802.03244 [hep-ph]}}.

\bibitem{Manohar:2016nzj}
A.~Manohar, P.~Nason, G.~P. Salam, and G.~Zanderighi, {\em {How bright is the
  proton? A precise determination of the photon parton distribution
  function}\/},  \href{http://dx.doi.org/10.1103/PhysRevLett.117.242002}{Phys.
  Rev. Lett. {\bf 117} (2016) no.~24, 242002},
\href{http://arxiv.org/abs/1607.04266}{{\tt arXiv:1607.04266 [hep-ph]}}.

\bibitem{Schmidt:2015zda}
C.~Schmidt, J.~Pumplin, D.~Stump, and C.~P. Yuan, {\em {CT14QED parton
  distribution functions from isolated photon production in deep inelastic
  scattering}\/},  \href{http://dx.doi.org/10.1103/PhysRevD.93.114015}{Phys.
  Rev. {\bf D93} (2016) no.~11, 114015},
\href{http://arxiv.org/abs/1509.02905}{{\tt arXiv:1509.02905 [hep-ph]}}.

\bibitem{Ball:2013hta}
{NNPDF} Collaboration, R.~D. Ball, V.~Bertone, S.~Carrazza, L.~Del~Debbio,
  S.~Forte, A.~Guffanti, N.~P. Hartland, and J.~Rojo, {\em {Parton
  distributions with QED corrections}\/},
  \href{http://dx.doi.org/10.1016/j.nuclphysb.2013.10.010}{Nucl. Phys. {\bf
  B877} (2013)  290--320},
\href{http://arxiv.org/abs/1308.0598}{{\tt arXiv:1308.0598 [hep-ph]}}.

\bibitem{Giuli:2017oii}
{xFitter Developers' Team} Collaboration, F.~Giuli et al., {\em {The photon PDF
  from high-mass Drell-Yan data at the LHC}\/},
  \href{http://dx.doi.org/10.1140/epjc/s10052-017-4931-5}{Eur. Phys. J. {\bf
  C77} (2017) no.~6, 400},
\href{http://arxiv.org/abs/1701.08553}{{\tt arXiv:1701.08553 [hep-ph]}}.

\bibitem{Aad:2015bwa}
{ATLAS} Collaboration, G.~Aad et al., {\em {Measurement of exclusive
  $\gamma\gamma\rightarrow \ell^+\ell^-$ production in proton-proton collisions
  at $\sqrt{s} = 7$ TeV with the ATLAS detector}\/},
  \href{http://dx.doi.org/10.1016/j.physletb.2015.07.069}{Phys. Lett. {\bf
  B749} (2015)  242--261},
\href{http://arxiv.org/abs/1506.07098}{{\tt arXiv:1506.07098 [hep-ex]}}.

\bibitem{Aaboud:2017oiq}
{ATLAS} Collaboration, M.~Aaboud et al., {\em {Measurement of the exclusive
  $\gamma \gamma \rightarrow \mu^+ \mu^-$ process in proton-proton collisions
  at $\sqrt{s}=13$ TeV with the ATLAS detector}\/},
  \href{http://dx.doi.org/10.1016/j.physletb.2017.12.043}{Phys. Lett. {\bf
  B777} (2018)  303--323},
\href{http://arxiv.org/abs/1708.04053}{{\tt arXiv:1708.04053 [hep-ex]}}.

\bibitem{Chatrchyan:2011ci}
{CMS} Collaboration, S.~Chatrchyan et al., {\em {Exclusive photon-photon
  production of muon pairs in proton-proton collisions at $\sqrt{s}=7$ TeV}\/},
   \href{http://dx.doi.org/10.1007/JHEP01(2012)052}{JHEP {\bf 01} (2012)  052},
\href{http://arxiv.org/abs/1111.5536}{{\tt arXiv:1111.5536 [hep-ex]}}.

\bibitem{Chatrchyan:2012tv}
{CMS} Collaboration, S.~Chatrchyan et al., {\em {Search for exclusive or
  semi-exclusive photon pair production and observation of exclusive and
  semi-exclusive electron pair production in $pp$ collisions at $\sqrt{s}=7$
  TeV}\/},  \href{http://dx.doi.org/10.1007/JHEP11(2012)080}{JHEP {\bf 11}
  (2012)  080},
\href{http://arxiv.org/abs/1209.1666}{{\tt arXiv:1209.1666 [hep-ex]}}.

\bibitem{Cms:2018het}
{CMS, TOTEM} Collaboration, A.~M. Sirunyan et al., {\em {Observation of
  proton-tagged, central (semi)exclusive production of high-mass lepton pairs
  in pp collisions at 13 TeV with the CMS-TOTEM precision proton
  spectrometer}\/},  \href{http://dx.doi.org/10.1007/JHEP07(2018)153}{JHEP {\bf
  07} (2018)  153},
\href{http://arxiv.org/abs/1803.04496}{{\tt arXiv:1803.04496 [hep-ex]}}.

\bibitem{Budnev:1974de}
V.~M. Budnev, I.~F. Ginzburg, G.~V. Meledin, and V.~G. Serbo, {\em {The two
  photon particle production mechanism. Physical problems. Applications.
  Equivalent photon approximation}\/},
\href{http://dx.doi.org/10.1016/0370-1573(75)90009-5}{Phys. Rept. {\bf 15}
  (1975)  181}.

\bibitem{Klein:2016yzr}
S.~R. Klein, J.~Nystrand, J.~Seger, Y.~Gorbunov, and J.~Butterworth, {\em
  {STARlight: A Monte Carlo simulation program for ultra-peripheral collisions
  of relativistic ions}\/},
  \href{http://dx.doi.org/10.1016/j.cpc.2016.10.016}{Comput. Phys. Commun. {\bf
  212} (2017)  258--268},
\href{http://arxiv.org/abs/1607.03838}{{\tt arXiv:1607.03838 [hep-ph]}}.

\bibitem{Gluck:2002fi}
M.~Gluck, C.~Pisano, and E.~Reya, {\em {The Polarized and unpolarized photon
  content of the nucleon}\/},
  \href{http://dx.doi.org/10.1016/S0370-2693(02)02125-1}{Phys. Lett. {\bf B540}
  (2002)  75--80},
\href{http://arxiv.org/abs/hep-ph/0206126}{{\tt arXiv:hep-ph/0206126
  [hep-ph]}}.

\bibitem{Martin:2004dh}
A.~D. Martin, R.~G. Roberts, W.~J. Stirling, and R.~S. Thorne, {\em {Parton
  distributions incorporating QED contributions}\/},
  \href{http://dx.doi.org/10.1140/epjc/s2004-02088-7}{Eur. Phys. J. {\bf C39}
  (2005)  155--161},
\href{http://arxiv.org/abs/hep-ph/0411040}{{\tt arXiv:hep-ph/0411040
  [hep-ph]}}.

\bibitem{Martin:2014nqa}
A.~D. Martin and M.~G. Ryskin, {\em {The photon PDF of the proton}\/},
  \href{http://dx.doi.org/10.1140/epjc/s10052-014-3040-y}{Eur. Phys. J. {\bf
  C74} (2014)  3040},
\href{http://arxiv.org/abs/1406.2118}{{\tt arXiv:1406.2118 [hep-ph]}}.

\bibitem{Schmidt:2014aba}
C.~Schmidt, J.~Pumplin, D.~Stump, and C.~P. Yuan, {\em {QED effects and Photon
  PDF in the CTEQ-TEA Global Analysis}\/},
\href{http://dx.doi.org/10.22323/1.203.0054}{PoS {\bf DIS2014} (2014)  054}.

\bibitem{Harland-Lang:2016kog}
L.~A. Harland-Lang, V.~A. Khoze, and M.~G. Ryskin, {\em {Photon-initiated
  processes at high mass}\/},
  \href{http://dx.doi.org/10.1103/PhysRevD.94.074008}{Phys. Rev. {\bf D94}
  (2016) no.~7, 074008},
\href{http://arxiv.org/abs/1607.04635}{{\tt arXiv:1607.04635 [hep-ph]}}.

\bibitem{Bertone:2017bme}
{NNPDF} Collaboration, V.~Bertone, S.~Carrazza, N.~P. Hartland, and J.~Rojo,
  {\em {Illuminating the photon content of the proton within a global PDF
  analysis}\/},  \href{http://dx.doi.org/10.21468/SciPostPhys.5.1.008}{SciPost
  Phys. {\bf 5} (2018)  008},
\href{http://arxiv.org/abs/1712.07053}{{\tt arXiv:1712.07053 [hep-ph]}}.

\bibitem{daSilveira:2014jla}
G.~G. da~Silveira, L.~Forthomme, K.~Piotrzkowski, W.~Schäfer, and A.~Szczurek,
  {\em {Central production via photon-photon fusion in proton-proton collisions
  with proton dissociation}\/},
  \href{http://dx.doi.org/10.1007/JHEP02(2015)159}{JHEP {\bf 02} (2015)  159},
\href{http://arxiv.org/abs/1409.1541}{{\tt arXiv:1409.1541 [hep-ph]}}.

\bibitem{Catani:1990eg}
S.~Catani, M.~Ciafaloni, and F.~Hautmann, {\em {High-energy factorization and
  small x heavy flavor production}\/},
\href{http://dx.doi.org/10.1016/0550-3213(91)90055-3}{Nucl. Phys. {\bf B366}
  (1991)  135--188}.

\bibitem{Aad:2008zzm}
{ATLAS} Collaboration, G.~Aad et al., {\em {The ATLAS Experiment at the CERN
  Large Hadron Collider}\/},
\href{http://dx.doi.org/10.1088/1748-0221/3/08/S08003}{JINST {\bf 3} (2008)
  S08003}.

\bibitem{Chatrchyan:2008aa}
{CMS} Collaboration, S.~Chatrchyan et al., {\em {The CMS Experiment at the CERN
  LHC}\/},
\href{http://dx.doi.org/10.1088/1748-0221/3/08/S08004}{JINST {\bf 3} (2008)
  S08004}.

\bibitem{Drell:1970wh}
S.~D. Drell and T.-M. Yan, {\em {Massive Lepton Pair Production in
  Hadron-Hadron Collisions at High-Energies}\/},
  \href{http://dx.doi.org/10.1103/PhysRevLett.25.316,
  10.1103/PhysRevLett.25.902.2}{Phys. Rev. Lett. {\bf 25} (1970)  316--320}.
[Erratum: Phys. Rev. Lett.25,902(1970)].

\bibitem{Aad:2015gta}
{ATLAS} Collaboration, G.~Aad et al., {\em {$Z$ boson production in $p+$Pb
  collisions at $\sqrt{s_{NN}}=5.02$ TeV measured with the ATLAS detector}\/},
  \href{http://dx.doi.org/10.1103/PhysRevC.92.044915}{Phys. Rev. {\bf C92}
  (2015) no.~4, 044915},
\href{http://arxiv.org/abs/1507.06232}{{\tt arXiv:1507.06232 [hep-ex]}}.

\bibitem{Khachatryan:2015pzs}
{CMS} Collaboration, V.~Khachatryan et al., {\em {Study of Z boson production
  in pPb collisions at $\sqrt {s_{NN}} = 5.02$ TeV}\/},
  \href{http://dx.doi.org/10.1016/j.physletb.2016.05.044}{Phys. Lett. {\bf
  B759} (2016)  36--57},
\href{http://arxiv.org/abs/1512.06461}{{\tt arXiv:1512.06461 [hep-ex]}}.

\bibitem{Alice:2016wka}
{ALICE} Collaboration, J.~Adam et al., {\em {W and Z boson production in p-Pb
  collisions at $\sqrt{s_{\rm NN}}$ = 5.02 TeV}\/},
  \href{http://dx.doi.org/10.1007/JHEP02(2017)077}{JHEP {\bf 02} (2017)  077},
\href{http://arxiv.org/abs/1611.03002}{{\tt arXiv:1611.03002 [nucl-ex]}}.

\bibitem{Dellacasa:1999ke}
{ALICE} Collaboration, G.~Dellacasa et al., {\em {ALICE technical design report
  of the zero degree calorimeter (ZDC)}\/}, .
CERN-LHCC-99-05.

\bibitem{ATLAS:2007aa}
{ATLAS Collaboration}, {\em {Zero degree calorimeters for ATLAS}\/}, .
CERN-LHCC-2007-01.

\bibitem{Manohar:2017eqh}
A.~V. Manohar, P.~Nason, G.~P. Salam, and G.~Zanderighi, {\em {The Photon
  Content of the Proton}\/},
  \href{http://dx.doi.org/10.1007/JHEP12(2017)046}{JHEP {\bf 12} (2017)  046},
\href{http://arxiv.org/abs/1708.01256}{{\tt arXiv:1708.01256 [hep-ph]}}.

\bibitem{Azevedo:2019fyz}
C.~Azevedo, V.~P. Goncalves, and B.~D. Moreira, {\em {Exclusive dilepton
  production in ultraperipheral $PbPb$ collisions at the LHC}\/},
  \href{http://dx.doi.org/10.1140/epjc/s10052-019-6952-8}{Eur. Phys. J. {\bf
  C79} (2019) no.~5, 432},
\href{http://arxiv.org/abs/1902.00268}{{\tt arXiv:1902.00268 [hep-ph]}}.

\bibitem{Abramowicz:1991xz}
H.~Abramowicz, E.~M. Levin, A.~Levy, and U.~Maor, {\em {A Parametrization of
  sigma-T (gamma* p) above the resonance region $Q^2 \geq 0$}\/},
\href{http://dx.doi.org/10.1016/0370-2693(91)90202-2}{Phys. Lett. {\bf B269}
  (1991)  465--476}.

\bibitem{Abramowicz:1997ms}
H.~Abramowicz and A.~Levy, {\em {The ALLM parameterization of sigma(tot)(gamma*
  p): An Update}\/},
\href{http://arxiv.org/abs/hep-ph/9712415}{{\tt arXiv:hep-ph/9712415
  [hep-ph]}}.

\bibitem{Suri:1971yx}
A.~Suri and D.~R. Yennie, {\em {The space-time phenomenology of photon
  absorption and inelastic electron scattering}\/},
\href{http://dx.doi.org/10.1016/0003-4916(72)90242-4}{Annals Phys. {\bf 72}
  (1972)  243}.

\bibitem{Vermaseren:1982cz}
J.~A.~M. Vermaseren, {\em {Two Photon Processes at Very High-Energies}\/},
\href{http://dx.doi.org/10.1016/0550-3213(83)90336-X}{Nucl. Phys. {\bf B229}
  (1983)  347--371}.

\bibitem{Szczurek:1999rd}
A.~Szczurek and V.~Uleshchenko, {\em {Nonpartonic components in the nucleon
  structure functions at small $Q^2$ in the broad range of x}\/},
  \href{http://dx.doi.org/10.1007/s100520000218}{Eur. Phys. J. {\bf C12} (2000)
   663--671},
\href{http://arxiv.org/abs/hep-ph/9904288}{{\tt arXiv:hep-ph/9904288
  [hep-ph]}}.

\bibitem{Martin:2009iq}
A.~D. Martin, W.~J. Stirling, R.~S. Thorne, and G.~Watt, {\em {Parton
  distributions for the LHC}\/},
  \href{http://dx.doi.org/10.1140/epjc/s10052-009-1072-5}{Eur. Phys. J. {\bf
  C63} (2009)  189--285},
\href{http://arxiv.org/abs/0901.0002}{{\tt arXiv:0901.0002 [hep-ph]}}.

\end{thebibliography}\endgroup

\end{document}